\documentclass{article}
\usepackage[utf8]{inputenc}

\usepackage{xr}
\usepackage{hyperref}
\usepackage{xr-hyper}
\usepackage{amsthm}
\usepackage{amsmath}
\usepackage{amssymb}
\usepackage{dsfont}
\usepackage{geometry}
\usepackage{booktabs}
\usepackage{graphicx}
\usepackage[shortlabels]{enumitem}
\usepackage{subcaption}
\usepackage{bm}
\usepackage{bbm}
\usepackage{url}
\usepackage{pifont}
\usepackage{verbatim}
\usepackage{listings}
\usepackage{dsfont}
\usepackage{nccmath}
\usepackage{tikz}
\usetikzlibrary{bayesnet}
\usepackage{authblk}

\usepackage{lineno}
\let\oldequation\equation
\let\oldendequation\endequation
\renewenvironment{equation}
  {\linenomathNonumbers\oldequation}
  {\oldendequation\endlinenomath}

\usepackage[
backend=bibtex,
style=authoryear,
citestyle=authoryear,
dashed=false,
maxcitenames = 2,
 mincitenames=1,
maxbibnames=1000,
natbib=true,
]{biblatex}
\bibliography{reference} 

\renewcommand{\d}{\mbox{d}}

\providecommand{\keywords}[1]
{
  \small	
  \textbf{\textit{Keywords: }} #1
}
\date{} % Comment this line to show today's date
\title{A hierarchical Bayesian non--asymptotic extreme value model for spatial data}
\author[1]{Federica Stolf}
\author[1]{Antonio Canale}
\affil[1]{Department of Statistical Sciences, University of Padova, Padova, Italy}
\affil[ ]{e-mail: federica.stolf@phd.unipd.it; antonio.canale@unipd.it}

\begin{document}
\maketitle

\begin{abstract}
Spatial maps of extreme precipitation are crucial in flood prevention.
With the aim of producing maps of precipitation return levels, we propose a novel approach to model a collection of spatially
distributed time series where the asymptotic assumption, typical of the traditional extreme value theory, is relaxed. We introduce a Bayesian hierarchical model that accounts for the possible underlying variability in the distribution of event magnitudes and occurrences, which are described through latent temporal and spatial processes. Spatial dependence is characterized by geographical covariates and effects not fully described by the covariates are captured by spatial structure in the hierarchies. The performance of the approach is illustrated through simulation studies and an application to daily rainfall extremes across North Carolina (USA). The results show that we significantly reduce the estimation uncertainty with respect to state of the art techniques.
\end{abstract} \hspace{10pt}

\keywords{Bayesian hierarchical models, Extreme value theory,  Spatial and temporal processes, Return levels, Rainfall extremes}
\maketitle

\section{Introduction}\label{sec1}

Extreme value theory deals with stochastic behavior of extreme events, found in the tails of probability distributions, and it finds wide application in environmental sciences. Events such as extreme precipitation and storm wind speed are driven by complex spatio--temporal processes and are usually characterized by limited predictability. Understanding frequency and intensity of these phenomena is important  in public--safety and long--term planning. 
Estimating the probability of extreme meteorological events is difficult because of limited temporal records and this issue is exacerbated in spatial settings, because forecasting entails the extrapolation in a large spatial domain. 

A variety of statistical tools have been used for modeling extreme value and the book by \citet{Coles2001} provides a gentle introduction to the topic.
Traditional methods are based on the block maxima approach exploiting the generalized extreme value distribution (GEV) or on the peaks over threshold (POT) approach exploiting the generalized Pareto distribution (GPD). 
The GEV is a three--parameter family of distributions that describes the asymptotic behavior of suitably renormalized block maxima of a sequence of independent and identically distributed random variables. The shape parameter of this distribution plays the crucial role of determining the weight of the upper tail of the density. 
The GPD, used under a POT approach, is a two--parameter family of distributions that is used to model excesses over a suitably chosen high threshold.  The GEV and the GPD distributions are deeply connected through a point process characterization \citep{smith1989, Davison.Smith1990} and in most practical applications they are routinely employed regardless the suitability of their asymptotic arguments.

In the spatial context several approaches for extreme values have been proposed in the literature.  Prediction at unobserved locations may be required, which can only be achieved with a proper spatial model.
A natural generalization of GEV and GPD distributions to a multivariate context are respectively the max-stable processes \citep{maxstable} and the $r$-Pareto processes \citep{ThibaudOptiz2015, deFondeville2018}.
The max-stability or threshold-stability properties provide a strong theoretical justification for these methods, which have been used in a wide variety of environmental applications. However, these asymptotic models have a strong tail dependence that persists at increasingly high levels, while data usually suggest that it should weaken instead \citep{huserw}. The max-stable and Pareto processes are always asymptotically dependent, i.e. extreme events have a positive probability of occurring simultaneously at distinct sites, no matter how extreme they are. This is a major limitation in practice, especially for environmental applications, where the spatial dependence tends to vanish for increasing quantile levels \citep{huserJASA}.  
A good revision of the classical models of spatial extreme is provided by \citet{padoan_davison}. They highlight three main classes of models: models built on max-stable processes \citep{maxstable, maxbrian}, copula models \citep{sgelf} and Bayesian hierarchical models \citep{cooley2007, sang_gelfand}. For a discussion about more recent progress in the modelling and inference for spatial extremes see \citet{huserw}.

To overcome the rigid asymptotic dependence structures of  the max-stable and Pareto processes, some contributions have focused on developing asymptotic independence and more flexible hybrid models that can bridge the two asymptotic dependence regimes \citep{huserJASA, bopp, OPITZ16}. Alternatively, the conditional spatial extreme approach aims at describing the spatial behavior of a random process conditional on single points being large  \citep{wadsworth, socean}.

Another limitation of GEV--based approaches is that the support of the GEV distribution depends on its parameter values. This feature can complicate inferential procedures, such as maximum likelihood estimation, and can be particularly problematic in a setting with multiple and interacting covariates. For example,  \citet{blendedGEV} propose to substitute the GEV distribution with the blended generalised extreme value (bGEV), which has the right tail of a Fr\'echet distribution and the left tail of a Gumbel distribution, resulting in a heavy--tailed distribution with a parameter free support. 

The wide popularity of the asymptotic methods led much of the extreme value literature to focus only on a small portion of data: the block maxima or few values above a threshold. The "ordinary" values from which the maxima are extracted are discarded, wasting much of the available information. In addition, in many environmental applications the number of yearly events is not sufficiently large for the asymptotic argument to hold as discussed, for example, by   \citet{Kou} and \citet{MARANI2015121} in hydrology. 
Another limitation of traditional extreme value theory is the assumption of a constant distribution for the ordinary events over time, since many phenomena display changes in the event magnitude generation process that are well established \citep{MARANI2015121}. 
\citet{hmev} introduced a Bayesian hierarchical model for extreme values building upon the so called metastatistical extreme value distribution (MEVD) \citep{MARANI2015121, marraMEVD, marra23}. This is a non--asymptotic extreme value approach in which a compound parametric distribution describes the entire range of ordinary values, with parameters varying across blocks. 

In this paper we introduce a spatial hierarchical Bayesian model to analyze extreme values of rainfall intensity. Bayesian hierarchical models for extremes value modelling represent an active area of research  \citep{ghoshmall, bracken18,hmev}. One of the main advantages of employing a Bayesian approach in this context is the possibility of incorporating expert domain knowledge in the form of informative prior distributions. This is dramatically relevant in a field like extreme value analysis where observations are by nature scarce and particularly in environmental studies where reliable expert prior information about the geophysical processes at hand is often available.
The proposed hierarchical Bayesian model focuses on modeling ordinary events and then estimating the cumulative distribution function (cdf) of maxima, making it intrinsically different from the models inspired by the GEV or max-stable processes. We refer to the proposed approach as a \emph{non--asymptotic} model, in the sense that it does not assume a divergent number of events in a block or use the approximations of the cdf of block maxima, as the GEV. 
Our proposal does not characterize extremal dependence at data level, so in the sense of  \citet{huserw} can be considered asymptotically independent, but it is capable of assessing marginal risks. 

In Section \ref{sec2} we introduce the general structure of the spatial hierarchical model and subsequently specialize it to the analysis of rainfall data. In Section \ref{sec3} the proposed formulation is tested through an extensive simulation study. An application to maximum rainfall data is described in Section \ref{sec4}. The paper ends with a final discussion. Code and data are available at \url{https://github.com/federicastolf/sHMEV}.

\section{A spatial Hierarchical Bayesian extreme value model}\label{sec2}

\subsection{Notation and general formulation} \label{sec:shmev_general}

Let $x_{ij}(s)$ denote the magnitude of the $i$-th event within the $j$-th block for site $s$, where $j = 1, \dots, J$ with $J$ the number of blocks in the observed sample , $i = 1, \dots, n_j(s)$ with $n_j(s)$ the number of events observed within the $j$-th block and $s \in \mathcal{A}$ with $\mathcal{A}$ a spatial domain. Let $S$ denote the total number of spatial points at which the data are observed within the spatial domain of interest $\mathcal{A}$, i.e. $S = |\cal{A} |$. 
Let  $Y_j(s) = \text{max}_{i} \{X_{ij}(s)\}$ denote the block maxima for each site $s$. As standard practice with environmental data, blocks have size of one year \citep{Coles2001}, thus by block maxima we refer to annual maxima. The main goal of extreme value analysis can be summarized in estimating the cumulative distribution function of  $Y_j(s)$.
In the classical extreme value framework, GEV distribution arises as the only possible limit model for block maxima. Specifically, if there exist sequences of constants $a_{j}(s) > 0$ and $b_{j}(s) \in \mathbb{R}$ such that, for $Y_j(s)^{*} = (Y_j(s) - b_{j}(s))/a_j(s)$, one has \mbox{$\mathrm{Pr}(Y_j(s)^{*} \le y) \rightarrow G(y)$} as $n_{j}(s) \rightarrow \infty$, where $G$ is a non--degenerate distribution  function, then G has the form
\begin{equation}\label{eq:gev}
G(y ; \mu, \sigma, \tau) = \exp\left\{- \left[ 1 + \tau \left( \frac{y-\mu}{\sigma} \right) \right]^{-1/\tau}_{+} \right\},
\end{equation}
with $a_{+} = \max(a,0)$, $\mu \in \mathbb{R}$ (location), $\sigma \in \mathbb{R}^{+}$ (scale) and $\tau \in  \mathbb{R}$ (shape), with support $\{y \in \mathbb{R} : 1+\tau(y-\mu) / \sigma >0\}$. This approach assumes that the limit GEV distribution is a good approximation for block maxima in finite samples.

Instead of relying on asymptotic arguments, we will exploit a fully generative hierarchical model. We focus on the ``ordinary'' values from which the maxima are extracted, in our application the daily records with non--zero rainfall,  and we model the magnitude $x_{ij}(s)$ and the occurrence $n_j(s)$ of these events. 
Notably, the proposed approach exploits the information contained in all the sample of ordinary values and not only the one contained in the block maxima, or in the values above a prespecified threshold, as done in the block maxima or peak over threshold approaches, respectively. 
 
 Specifically, we assume that $x_{ij}(s)$, conditionally on unobserved latent processes, are realizations of conditionally independent random variables $X_{ij}(s)$. We further assume that each $X_{ij}(s)$ has a location and time--specific distribution, but with all distributions belonging to the same parametric family with cdf $F(\cdot;\theta_j(s))$, with $\theta_j(s) \in \Theta$ unknown parameter vector of suitable dimension. We denote with {$f(\cdot; \theta_j(s))$ the related probability density function. The main goal is to estimate the cdf of the block maxima:
 \begin{equation} \label{eq:cdf_max}
\mbox{Pr}(Y_j(s) \le y) = F(y; \theta_{j}(s))^{n_j(s)}.
\end{equation}
 The following section specifies this general notation focusing on the case when  $x_{ij}(s)$ are daily rainfall accumulations. 

We assume that the response variables are influenced by both temporal and spatial underlying processes. The base layer models the magnitude of events at each spatial location for each block. The nested layer subsequently models the latent processes that drive the parameters of the external layer, both in time and space. Using a Bayesian approach, the last layer is associated with the prior distributions of the unknown parameters that govern the inner latent processes. The structure of the model is illustrated in Figure \ref{fig:shmver}.

Let $n_j(s)$ be a realization of a stochastic process with conditional probability function $p\{\cdot;\lambda(s)\}$, where $\lambda(s)$ is an unknown parameter vector depending on the spatial location $s\in \cal A$.
We further assume that  $\theta_j(s)$ are realizations of a stochastic process with conditional probability density function $g\{\cdot ;\eta(s)\}$, where $\eta(s)$ is an unknown vector of parameters.  
The proposed Bayesian spatial Hierarchical Model for Extreme Values (sHMEV) can be written in hierarchical form, for each $s\in \cal A$, as 

\begin{equation}
  \label{eq:shmev_general}
  \begin{aligned}
   & x_{ij}(s)  |n_j(s),\theta_j(s)  \sim f\{x_{ij}(s);\theta_j(s)\},\,\, \mbox{for $i=1, \dots, n_j(s)$},\\        
     &\theta_j(s)  | \eta(s)  \sim g\{\theta_j(s);\eta(s)\}, \quad n_j(s)|\lambda(s) \sim p\{n_j(s); \lambda(s)\}. \\
  \end{aligned}
\end{equation}
The latent spatial processes, $\eta(s)$ and $\lambda(s)$, can be driven by unknown parameters $\beta_\eta$ and $\beta_\lambda$ in a stochastic manner as 
\begin{equation}
  \label{eq:shmev_general_2}
  \begin{aligned}
   & \lambda(s)  |\beta_\lambda  \sim k\{\lambda(s);\beta_\lambda\}, \quad \eta(s)|\beta_\eta \sim m\{\eta(s);\beta_\eta\},
  \end{aligned}
\end{equation}
where $k(\cdot; \beta_\lambda)$ and $m(\cdot;\beta_\eta)$ are suitable probability density functions. A simplified model assumes for $\eta(s)$ and $\lambda(s)$ simple deterministic functions of parameters $\beta_\eta$ and $\beta_\lambda$ and spatial covariates $z(s)$. The Bayesian representation of the model is completed by eliciting suitable prior distributions for the unknown parameters $\beta_\eta$ and $\beta_\lambda$.

Having as target the cdf in (\ref{eq:cdf_max}) or one of its functionals, the estimation can be done in our setting marginalizing  out the variables $n_j(s)$ and $\theta_j(s)$, i.e.
\begin{equation} \label{eq:h}
    \zeta\{y;\lambda(s),\eta(s)\} = \sum_{n \ge 0} \int_{\Theta} F\{y;\theta\}^{n} \,p\{n;\lambda(s)\} \, g\{\theta;\eta(s)\}  \d\theta.
\end{equation}

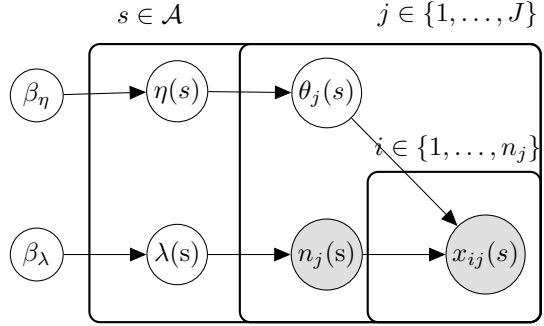
\begin{figure} [t]
\centering
\begin{tikzpicture}
%node
\node[latent](b){$\beta_{\lambda}$};
\node[latent, right=1.1 of b](l){$\lambda$(s)}; 
\node[obs, right=1.1 of l](n){$n_j$(s)}; 
\node[obs, right=1.1 of n](xij){$x_{ij}(s)$}; 
\node [latent, above=1.4 of b] (a) {$\beta_{\eta}$};
\node [latent, above=1.35 of l] (eta) {$\eta(s)$};
\node [latent, above=1.2 of n] (th) {$\theta_j(s)$};

\edge{b}{l}
\edge{l}{n}
\edge{a}{eta}
\edge{eta}{th}
\edge{th,n}{xij}

\tikzset{rounded_box/.style={draw, inner sep=2mm, rectangle, rounded corners}};
\draw[thick, rounded corners] (4.4,-0.9) rectangle (6.7,1.1);
\draw[thick, rounded corners] (2.7,-0.9) rectangle (6.7,2.8);
\draw[thick, rounded corners] (0.7,-0.9) rectangle (6.7,2.8);
\node at (5.6,3.2) { $j \in \{1,\dots,J\}$};
\node at (5.6,1.4) { $i \in \{1,\dots,n_j\}$};
\node at (1.5,3.2) { $s \in \cal{A}$};

\end{tikzpicture}
\caption{Graphical representation of the spatial hierarchical model described in (\ref{eq:shmev_general}) and (\ref{eq:shmev_general_2}). }
\label{fig:shmver}
\end{figure}

\subsection{A specific formulation for modelling daily rainfall} \label{sHMEV_rain}

Hereafter we specify the approach described in the previous section to the case of annual maxima daily rainfall accumulations over an area of interest. Several parametric families have been employed to model rainfall accumulations and while most of the approaches are based on goodness of fit considerations, ours exploits physical knowledge of the phenomena. It is well established that precipitation is a heavy--tailed phenomenon \citep{katz} and \citet{stechmann} suggest that the distribution of daily rainfall should follow a gamma distribution, while \citet{wilsonw} and  \citet{porporato2006}  suggest that its right tail should decay as a stretched exponential (i.e., Weibull) distribution.
Consistently with these arguments, we believe that the Weibull distribution is adequate for modeling the magnitudes of daily rainfall accumulations. In other applications, however, it is important to exercise prudence in selecting the parent distribution for event magnitudes, as this constitutes a critical component of our methodology. It is imperative to note that inaccuracies in this regard could potentially have an adverse impact on the final results.

Thus, we assume that $x_{ij}(s)$ (for $i=1, \dots, n_{j}(s), \ j=1,\dots,J \text{ and } s=1, \dots, S)$ follow a Weibull distribution with parameter vector $\theta_j(s) = (\gamma_j(s), \delta_j(s))$, where $\delta_j(s)>0$ denotes the scale parameter and  $\gamma_j(s) > 0$ the shape parameter. To allow for the inter-block variability we assume that the latent variables $\gamma_j(s) \sim g_{\gamma}\{\gamma_j(s);\mu_{\gamma}(s), \sigma_{\gamma}\}$ and $\delta_j(s) \sim g_{\delta}\{\delta_j(s);\mu_{\delta}(s), \sigma_{\delta}\}$ are independent and follow a Gumbel distribution, a flexible yet parsimonious model allowing for possible asymmetry. We further assume that only the location parameters of the two Gumbel distributions are characterized by a spatial dependence and not their scale parameters, but more flexible alternatives are straightforward. 
For the location processes $\mu_{\gamma}(s)$ and $\mu_{\delta}(s)$ we assume a linear dependence from fixed spatial covariates $z(s)$, i.e.
\begin{equation}
   \mu_{\gamma}(s) =z(s) \beta_{\gamma}, \quad    \mu_{\delta}(s) =z(s) \beta_{\delta} 
   \label{eq:regress}
\end{equation}
where  $\bm{\beta_{\gamma}} = \big[\begin{matrix}
\beta_{\gamma,0} & \beta_{\gamma,1} & \cdots & \beta_{\gamma,p}
\end{matrix}\big]^T $ and $\bm{\beta_{\delta}} = \big[\begin{matrix}
\beta_{\delta,0} & \beta_{\delta,1} & \cdots & \beta_{\delta,p}
\end{matrix}\big]^T$ and $p$ is the number of available spatial covariates. The vectors $\bm{\beta_{\gamma}}$ and $\bm{\beta_{\delta}}$ are independent and, in principle, we could also assume that they are of different size and related to different subsets of variables in $z(s)$. 
Although we focus on simple linear relations between the location processes and the spatial covariates, extensions to more flexible modelling structures are also straightforward, e.g. adopting basis expansions.
The idea underlying this specification is that observations are characterized by a latent process defined by geographic covariates, while the residual spatial variability, not captured by covariates, is described by the random Gumbel variability.

For $n_j(s)$ we assume a binomial distribution, with a success probability $\lambda(s)$ and number of trials $N_t$ equal to the block size (e.g., $N_t = 366$ in our application to annual maximum daily rainfall). The decision to employ a simple parametric model is supported by the consideration that the distribution of $n_j$ mainly affects the distribution of extreme events only through its average value \citep{marra19, hosseini20}. We assume that the rainfall occurrence is also affected by the geographical characteristics of the sites, as previously done for the location parameters of the Gumbel distributions. Consistently with (\ref{eq:regress}) we let 
\begin{equation}
   \text{logit}( \lambda(s)) = z(s) \bm{\beta_{\lambda}}, 
\label{eq:logitlambda}
\end{equation}
where the function $\text{logit}(x) = \text{log}(x/(1-x))$ is employed for ensure $\lambda(s) \in (0,1)$. 
Specific examples of (\ref{eq:regress}) and (\ref{eq:logitlambda})  are discussed in Section \ref{sec4}.

We have assumed that the distributions of $n_{j}(s)$ and $x_{ij}(s)$ are independent, although one might expect a possible dependence between the number of the events that occur within a block and their distribution.
Indeed, precipitation is composed of a mixture of high-intensity events that occur over short-time scales and 
low--intensity, long--time scale events, and there may be a negative relationship between the occurrence and the event distribution. It is important to note, however, that our data are daily and hence we are summing up all the short-term events occurring on the same day, thus partially losing the details of the most intense convective rainfall.

\subsection{Prior elicitation and posterior computation} \label{sec:prior}

The introduction of a hierarchical model that describes the entire distribution of daily rainfall accumulations allows to specify priors directly on the underlying distribution of the observed ordinary events, rather than on the distribution of block maxima. This is a great advantage, because it avoids the difficulty of prescribing a prior directly on the shape parameter $\zeta$ of the GEV distribution, to which it is difficult to attribute physical meaning.

In defining the prior distributions for the parameters $\sigma_{\gamma}$, $\sigma_{\delta}$, $\bm{\beta_{\delta}}$, $\bm{\beta_{\gamma}}$, and $\bm{\beta_{\lambda}}$, we seek to harness information on the physical processes generating the data, avoiding, where possible, strongly uninformative priors. For example, \citet{sornette, Frisch_1997} provide geophysical motivations to expect the Weibull shape parameters $\gamma(s)$ to be centered around 2/3  for rain accumulation. Consistently with this, we fix the prior expectation of $\beta_{\gamma,0}$ to $2/3$. 
In the present study, selecting a highly informative prior for $\beta_{\gamma,0}$ was predicated on a sound basis of expert domain knowledge. However, it is worth noting that such a justification may not be applicable in all contexts. In recognition of this potential limitation, we have undertaken a sensitivity analysis to examine the robustness of the posterior distribution of $\beta_{\gamma,0}$ in relation to the choice of hyperparameters for the prior. The analysis showed that results are robust to moderate and reasonable changes in the hyperparameter of the prior for $\beta_{\gamma,0}$.
The remaining terms in the vector $\beta_\gamma$ are assumed having null prior expectations, as the covariates have been standardized. 
As parametric distribution for each vector $\bm{\beta_{\delta}}$, $\bm{\beta_{\gamma}}$ and  $\bm{\beta_{\lambda}}$  we chose  independent Gaussian distributions.
For the latent Gumbel scale parameters $\sigma_{\gamma}$ and $\sigma_{\delta}$, quantifying the between--block  variability of the Weibull parameters, we opt for independent inverse gamma distributions, with expectations equal to 25\% and 5\% of the respective location parameters ($\mu_{\delta}$ and $\mu_{\gamma}$). This choice reflects the expectation that the scale parameter varies across years more than the shape parameter, as its expected values should be more constrained by the behavior of the physical process of precipitation.
For the choice of the hyperparameters of the distributions we adopt an empirical Bayesian approach \citep{Casella1985AnIT} and a practical example is illustrated in Section \ref{sec4}.

Given the complex structure of the introduced model,  the posterior distribution of the parameters is not available in closed form. Posterior approximation is then obtained using Markov Chain Monte Carlo (MCMC) methods. Specifically, we use the \textit{Hamiltonian Monte Carlo} (HMC) approach \citep{betancourt2018conceptual} exploiting the flexibility of the Stan software \citep{carpenter2017}.

In all the following examples, we run $n_c$ = 4 parallel chains, with $n_g$ = 2000 iterations in each chain, starting from different initial points. We discard the first half of each chain to account for the burn-in effect. Therefore, the final sample on which we perform inference is based on $B = n_c n_g/2 = 4000$ draws, obtained by merging the draws from different chains, following standard practice \citep{gelmanbda03}.

We can obtain an estimation of (\ref{eq:h}), the cumulative probability of block maxima approximating at each location $s \in \cal A$, with a two step procedure. We first compute the value of $\hat{\zeta}_s^{(b)}(y)$ at the generic $b$ iteration for  site $s$ as
\begin{equation} \label{quant_sHMEV2}
    \hat{\zeta}^{(b)}_{s}(y) = \frac{1}{M_g} \sum_{j=1}^{M_g} F\{y;\theta_{j}^{(b)}(s)\}^{n_{j}^{(b)}(s)},
\end{equation}
where $\theta_{j}^{(b)}(s) = \big(\gamma_{j}^{(b)}(s), \delta_{j}^{(b)}(s)\big)$ and $n_{j}^{(b)}(s)$, for $j = 1, \dots, M_g$, are drawn from the related posterior predictive distribution for site $s$ in the $j$-th over $M_g$ future blocks. Then, averaging over the $B$ draws from the posterior distribution for each site $s$ we have 
\begin{equation} \label{quant_sHMEV1}
   \hat{\zeta}_{s}(y) = \frac{1}{B} \sum_{b=1}^{B} \hat{\zeta}^{(b)}_{s}(y).
\end{equation}

\section{Simulation study}\label{sec3}

\subsection{Description}

To assess the empirical performance of the proposed model we perform a simulation study. A data set composed of $S = 27$ sites has been constructed, simulating uniformly inside a square of unitary side. Taking as reference point the bottom left edge of the square, the values on the x-axis and the y-axis represent the spatial coordinates, $z_1$ and $z_2$.

We  simulated the ordinary events and extract the annual maxima from them, since we believe that this would be more realistic than simulating the maxima directly from an asymptotic approximation, like the GEV distribution. Different synthetic data sets have been generated under three scenarios characterized by specific event magnitude distributions.
In the first scenario (WEI) we assume a Weibull model in which the scale and shape parameters in each block follow Gumbel distributions. The location parameters of the Gumbel distributions are determined through a spatial trend, defined  as:
\begin{equation} \label{trend_sim}
   t(s) =  \beta_0 + \beta_1 \ z_1(s) + \beta_2 \ z_2(s),
\end{equation}
with $\beta_0, \beta_1, \beta_2 \in \mathbb{R}$ and $s \in  \cal{A}$. 
In the second scenario (WEI$_{gp}$) the model builds upon the previous one and in order to give some extra variability to the data, we add a Gaussian process with exponential correlation function to the regression trend defined in (\ref{trend_sim}). Thus the location parameters of the Gumbel distributions are defined through the function
\begin{equation} \label{trend_sim_gp}
   t(s) =  \beta_0 + \beta_1 \ z_1(s) + \beta_2 \ z_2(s) + \mathcal{GP} (s; \alpha, \nu ),
\end{equation}
where $\cal{GP}$ is a zero mean Gaussian process with covariance function $\alpha \exp(- ||d || / \nu)$ with parameters $\alpha$ and $\nu$, while  $d$ denotes the pairwise Euclidean distance between locations.
In the third specification (GM) we assume for the event magnitude a Gamma distribution, where the two parameters are generated by a spatial trend defined accordingly to Equation (\ref{trend_sim}).

The number of events in each block is drawn from a binomial distribution with number of trials $N_t = 366$ and success probability determined through a logit transformation of the spatial trend defined in (\ref{trend_sim}) with constant value of $\beta_0, \beta_1,$ and $\beta_2$ for all  the three scenarios.
Notably, only the first scenario reflects the structure of the proposed spatial hierarchical model. This will be used  to assess the robustness of the proposed formulation under model misspecification.

Common to all scenarios, two independent data sets have been generated under the same specifications: the first one, which contains time series of length 20 years, will be used as the training set, while the second one, containing time series of length 100 years, will be used as the test set. Using limited temporal records for the estimation is representative of many real geophysical data sets.

The sHMEV model assumes, under correct specification of (\ref{trend_sim}), 
\begin{equation*}
\begin{gathered}
   \mu_{\gamma}(s) = \beta_{\gamma,0} + \beta_{\gamma,1} z_1(s) + \beta_{\gamma,2} z_2(s), \quad 
      \mu_{\delta}(s) = \beta_{\delta,0} + \beta_{\delta,1} z_1(s) + \beta_{\delta,2} z_2(s),\\
      \mbox{logit}(\lambda(s)) = \beta_{\lambda,0} + \beta_{\lambda,1} z_1(s) + \beta_{\lambda,2} z_2(s), 
      \end{gathered}
\end{equation*}

where $z_1(s)$ and $z_2(s)$ are the spatial coordinates for site $s$, with $s \in \cal{A}$.

\subsection{Performance measures}
To evaluate the performance of the proposed spatial hierarchical model we compare it with standard alternative methods. In particular, the methods used to benchmark sHMEV are Bayesian implementations of the classical generalized extreme value distribution (GEV) and the Bayesian hierarchical model (HMEV)  described in \citet{hmev}. Between the competing models we did not include the peaks over threshold approach because the predictive results of the latter model were similar to those of GEV.

For both the models used to benchmark sHMEV we take informative priors. More precisely, for the HMEV model we follow the specification of \citet{hmev}, while for the GEV model the prior distribution for the shape parameter is centered around the value 0.114 and has a standard deviation of 0.125; these values have been determined from investigations of rainfall records at the global scale \citep{papax}.

To compare the different methods, we evaluate the predictive accuracy in estimating the distribution of block maxima on the test set. We introduce different criteria to measure the predictive performance of the methods for quantiles above a given non exceedance probability, that are computed marginally for each station.

Specifically, we consider the fractional squared error (FSE) defined by
\begin{equation}
     \frac{1}{m_T} \sum_{j=1}^{M_x} \mathds{1}_{(\Tilde{T},\infty)} \{T_{js}\} \sqrt{\frac{1}{B} \sum_{b=1}^B \bigg(\frac{\hat{\zeta}_s^{(b)^{-1}}(p_{js})-y_{js}}{y_{js}}\bigg)^2},
\end{equation}
where $\mathds{1}_A(x)$ is the indicator function that equals 1 if x belongs to A, 
$\hat{\zeta}_s^{(b)^{-1}}(\cdot)$ is  the quantile function of the specific model at the $b$-th MCMC iteration for site $s$, 
$T_{js}$ is the empirical return time of $y_{js}$  defined as $ T_{js} = (1 - p_{js})^{-1}$, with $p_{js} = \text{rank}(y_{js})/(M_x + 1)$ and $M_x$ is the number of blocks used to compute the FSE. In this section $M_x$ corresponds to the number of blocks in the test set, i.e.  $M_x = 100$. The value $m_T$ represents the number of observations in the test set with empirical return time equal to or larger than $\Tilde{T}$, i.e. $m_T = \sum_{j=1}^{M_x} \mathds{1}_{(\Tilde{T},\infty)} (T_{js})$. 
The FSE index represents an average measure of a standardized distance between model-estimated quantiles and empirical quantiles for return times larger than $\Tilde{T}$.
In the following analysis we define $\Tilde{T} = 2$, consistent with  the range of exceedance probability of interest in many practical applications.

To separately assess the precision and the variability of extreme value quantile estimates, we employ two additional measures, namely the average bias and the average width of the $ 90\% $ posterior predictive credible intervals, defined respectively as
\begin{equation} \label{eq:mbias}
    b_q = \frac{1}{m_T} \sum_{j=1}^{M_x} \mathds{1}_{(\Tilde{T},\infty)} \{ T_{js}\} \frac{1}{B} \sum_{b=1}^B \bigg(\frac{\hat{\zeta}_s^{(b)^{-1}}(p_{js})-y_{js}}{y_{js}}\bigg),
\end{equation}
\begin{equation} \label{eq:mwidth}
    \Delta_{q90} = \frac{1}{m_T} \sum_{j=1}^{M_x} \mathds{1}_{(\Tilde{T},\infty)} (T_{js}) (\hat{q}_{95}(p_{js}) - \hat{q}_5(p_{js})),
\end{equation}
where the quantities $\hat{q}_{95}(p_{js})$ and $\hat{q}_{5}(p_{js})$ are the upper and lower bounds of the posterior credibility interval for the quantile $\hat{\zeta}_s^{(b)^{-1}}(p_{js})$.

\subsection{Results}
Figure \ref{fig:fse_sim} shows the empirical distribution of the FSE over the sites, computed on the test set. The proposed sHMEV outperforms the competitors in all scenarios, even if we can see that in the GM scenario the interquartile variability of the FSE index for different sites is higher with respect to the other two models. In the WEI$_{gp}$ scenario the GEV model has some rather high FSE values, probably a consequence of  a consistent inter--block  variability for the distribution of $x_{ij}(s)$.

To gain a deeper understanding of this general behavior, Figure~\ref{fig:mbmw_sim} reports the results of the two criteria introduced in (\ref{eq:mbias}) and (\ref{eq:mwidth}).
Generally, the best performance for the bias appears to be specification dependent. In the WEI and GM scenarios, sHMEV tends to slightly underestimate the posterior predictive quantiles and HMEV gets the best results. In  the WEI$_{gp}$ scenario, instead, sHMEV shows the lower bias.  For the width of the credibility intervals, sHMEV is consistently the most efficient procedure, producing narrower credibility intervals. 
These considerations suggest that the variability of the estimates affects the calculation of the FSE index more than the bias, since the latter is reduced for all three competitors.

\begin{figure}[t]
	\centering
	\includegraphics[height = 0.22\textheight]{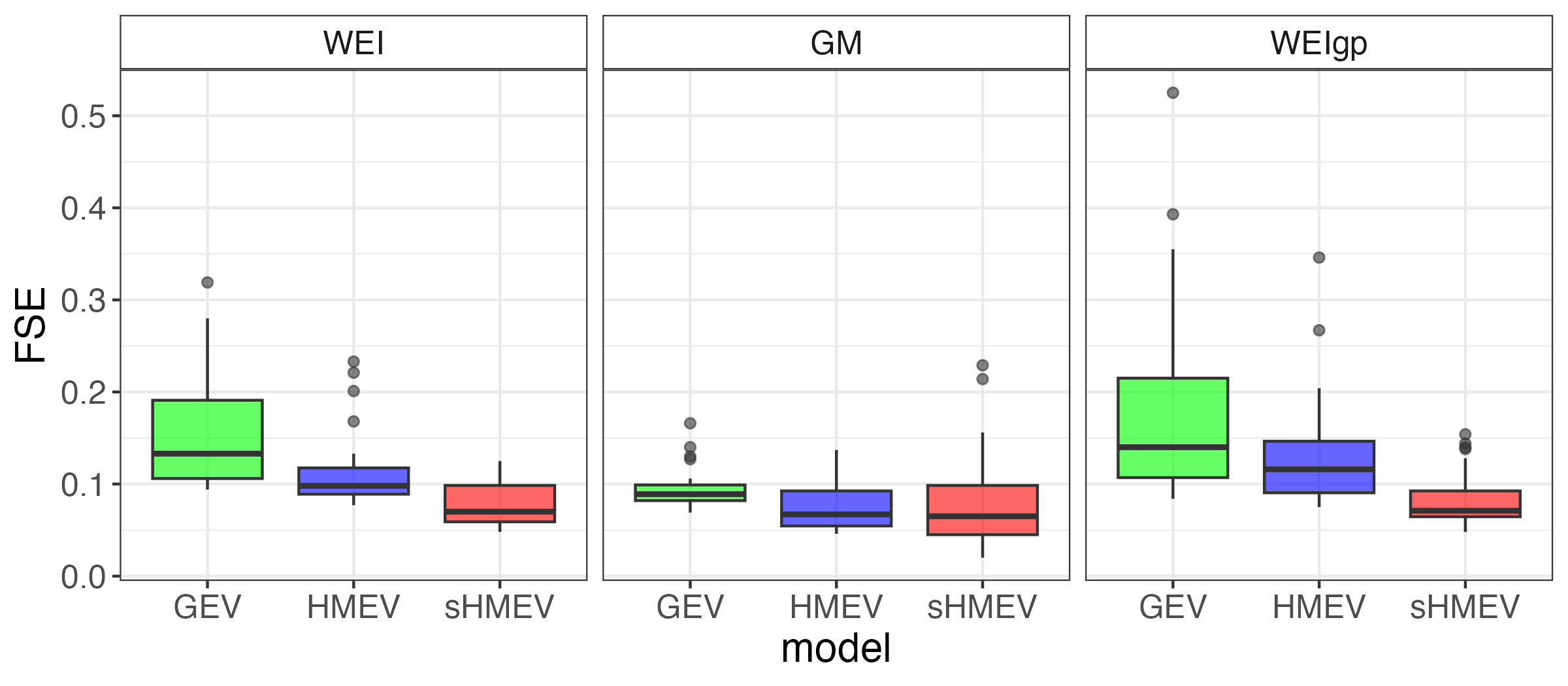}
	\caption{Fractional square error computed for the 3 different model specifications.}  
	\label{fig:fse_sim}
\end{figure}

\begin{figure}[h]
	\centering
	\includegraphics[height = 0.22\textheight]{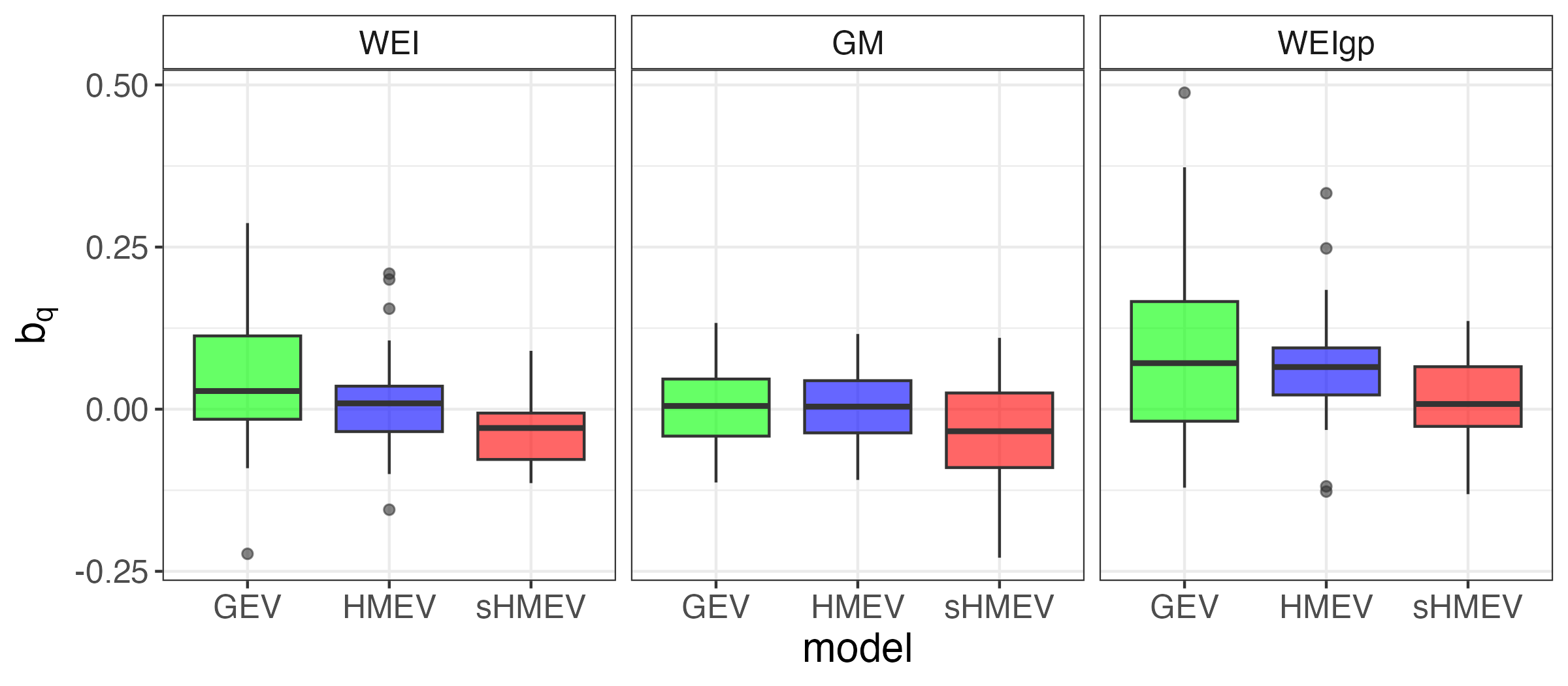}
	\includegraphics[height = 0.22\textheight]{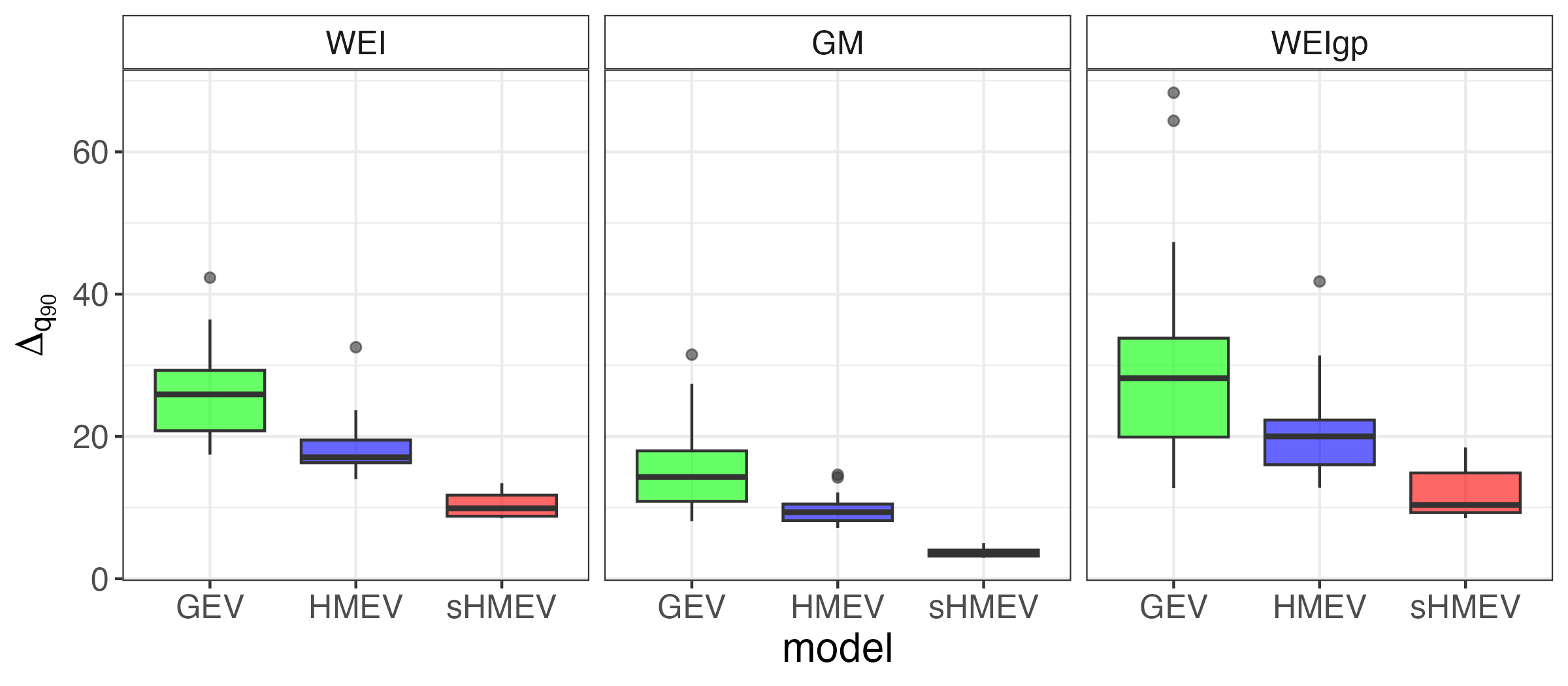}
	\caption{Mean bias (top row) and mean credibility interval width (bottom row) computed for the 3 different model specifications.}  
	\label{fig:mbmw_sim}
\end{figure} 
\begin{figure}[h!]
	\centering
	\includegraphics[width=0.63\textwidth]{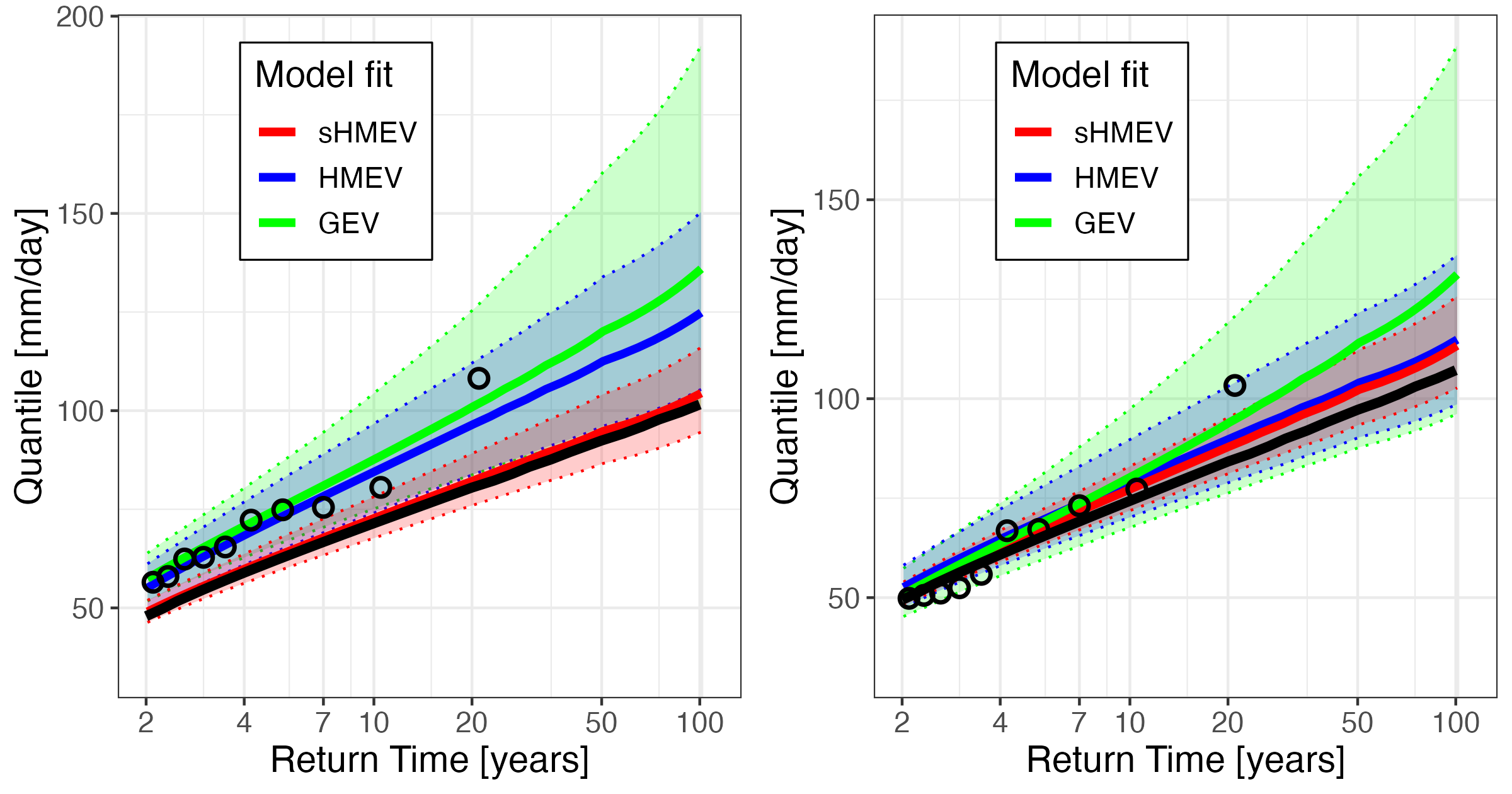}
	\caption{Quantiles predicted for two sites by the GEV (green), HMEV (blue), and sHMEV (red) models based on data simulated in the first scenario (WEI). Solid lines show the expected value of the quantile for a given return time, while dashed lines represent the bounds of 90\% credibility intervals.  Circles represent the observed block maxima on the training set, while the black lines report the quantiles computed from the true sHMEV model.}  
	\label{fig:quant_sim}
\end{figure} 

To visualize this global behavior, Figure \ref{fig:quant_sim} shows a representative example of the performance of the methods. Specifically, it reports the quantile versus return time plots obtained for the different methods applied to the data generated under hypothesis of correct model specification (WEI) for two randomly selected sites. The plots for all the remaining sites are shown in Figure S1--S2 in the supplementary materials.  
The quantiles computed from the true model are overall satisfactorily captured by the proposed model.
The sHMEV model yields quantile estimates with narrower credibility intervals, especially compared to the GEV model, and it is characterized by lighter tails than the other two methods.
Note that the GEV model, despite the informative prior used, appears to be more sensitive to the largest observations in the training set and tends to overestimate the true quantile function, as noted for the site in the right panel of Figure \ref{fig:quant_sim}.
This behavior is expected, given the limited length of the training samples used here, and consistent with previous studies \citep{hmev}.
For the site considered in the left panel of Figure \ref{fig:quant_sim} the maxima observed on the training set are quite different from the expected values on the test set. This situation depicts a common challenge in extreme value analysis, where limited temporal records result in biased predictions due to insufficient information.
In this case sHMEV, also exploiting information from the other sites (\textit{borrowing strength}), obtains more accurate and less variable estimations than the competitors. 

\section{Application: rainfall in North Carolina}\label{sec4}

\subsection{United States Historical Climatological Network data}
The data analyzed in this section are extracted from the United States Historical Climatological Network (USHCN) data, that are freely available from the National Centers for Environmental Information (NCEI) of the National Oceanic and Atmospheric Administration (NOAA) \citep{nooa}. The data consist of daily precipitation records for all the available weather stations in North Carolina, for the time period 1870 through 2021, with a significant fraction of the available records being longer than 100 years. The region is characterized by heterogeneity in morphological and climatic features: it varies from the plain areas near the coastline, to the hilly and mountainous zones in the west of the region. This allows us to test the proposed model in different climates and precipitation regimes.

The records characterized by non-blank quality flags were removed, as well as the years characterized by more than 30 daily missing observations. Therefore, for the subsequent analysis we select only stations that contain more than 73 years of data, for a total of 27 stations. We randomly chose 25 stations to fit our model and for each one we take the first 20 years. 

Figure \ref{fig:stazioni_nc} shows the station locations, with black points and triangles indicating stations included in the training set and blue points indicating stations included in the test set. Black triangles indicate three stations that will be used for illustrative purposes in the following.
One of these three stations is on the coastline (Edenton), another one is in the middle of the region (Fayetteville) and the last one is in the mountainous area (Hendersonville).

To assess the temporal dependence, we plot the autocorrelations of the daily rainfall accumulations time series observed at each station. 
Figure S3 in the supplementary materials shows the ACF plots for the three stations taken as examples, but the trend is the same for all stations considered in the analysis. 
As the locations fail to show any significant temporal dependence, operations to enhance the events pseudo-independent (e.g. declustering) are not necessary.

We tested for spatial dependence in the positive daily precipitation residuals using an exponential variogram with a linear trend and found a low dependence between stations within 20 km. Since the two closest stations are 24 km apart, the choice to model the spatial dependence with a latent process driven by spatial covariates seems appropriate.

Figure S4, in the supplementary materials, depicts the annual maxima of precipitation for all stations, where the color and the size of the circles are respectively proportional to the mean and the standard error of the annual maxima. The greatest intensity and variability of the maxima occur for stations on the coast and for some in the mountainous area. Therefore, we decided to consider as geographical covariates to be included in the model, in addition to latitude and longitude, also the altitude and the distance from the coast in km. These, in fact, appear to have a marginal effect on both the two parameters of the Weibull distribution, scale and shape, estimated via the method of moments for positive accumulations and on the number of annual rainy days.

\subsection{Fit of the sHMEV model}

We apply to the data the sHMEV model specified in Section \ref{sHMEV_rain}, using for the estimation the first 20 years of 25 stations, which means $J = 20$ and $S=25$.
According to the considerations made in the previous section, we define $\mu_{\gamma}(s)$ as

\begin{equation}
   \mu_{\gamma}(s) = \beta_{\gamma,0} + \beta_{\gamma,1} \text{lat}(s) + \beta_{\gamma,2} \text{lon}(s) + \beta_{\gamma,3} \text{alt}(s) + \beta_{\gamma,4} \text{dist}(s),
\end{equation}
where lat($s$), lon($s$), alt($s$) and dist($s$) are respectively latitude, longitude, altitude and distance from the coast of site $s$. All the covariates have been standardized. The functions for the parameters $\delta$ and $\lambda$ are similarly defined. We focus on a simple specification, since more flexible formulations that include Gaussian processes to model additional unexplained heterogeneity had obtained worse predictive performance. 

For prior elicitation, we follow what is reported in Section \ref{sec:prior}, avoiding the use of particularly uninformative distributions. We describe below the procedure followed for choosing the hyperparameters of the normal distributions for $\bm{\beta_{\delta}}$, but similar reasoning was used for the selection of the other hyperparameters. Recalling that the covariates were standardized, the value of the intercept $\beta_{\delta,0}$ refers to the average case, that is the case in which all predictors take on average value. For this parameter we define a prior distribution centered on the mean for the 25 stations of the parameter $\delta$ estimated on the data by the method of moments. The variance of the distribution is chosen such that the probability that $\beta_{\delta,0}$ is between two values considered reasonable is greater than 0.95. For the parameter $\beta_{\delta,1}$ we take a prior distribution centered on the least squares estimation of the simple regression of $\delta$ (always estimated for the 25 stations by the method of moments) on latitude. The variance was chosen equivalently to what was done for $\beta_{\delta,0}$. We follow a similar procedure for $\beta_{\delta,2}$, $\beta_{\delta,3}$ and  $\beta_{\delta,4}$.

\begin{figure}[t]  
	\centering
		\includegraphics[width=0.7\textwidth]{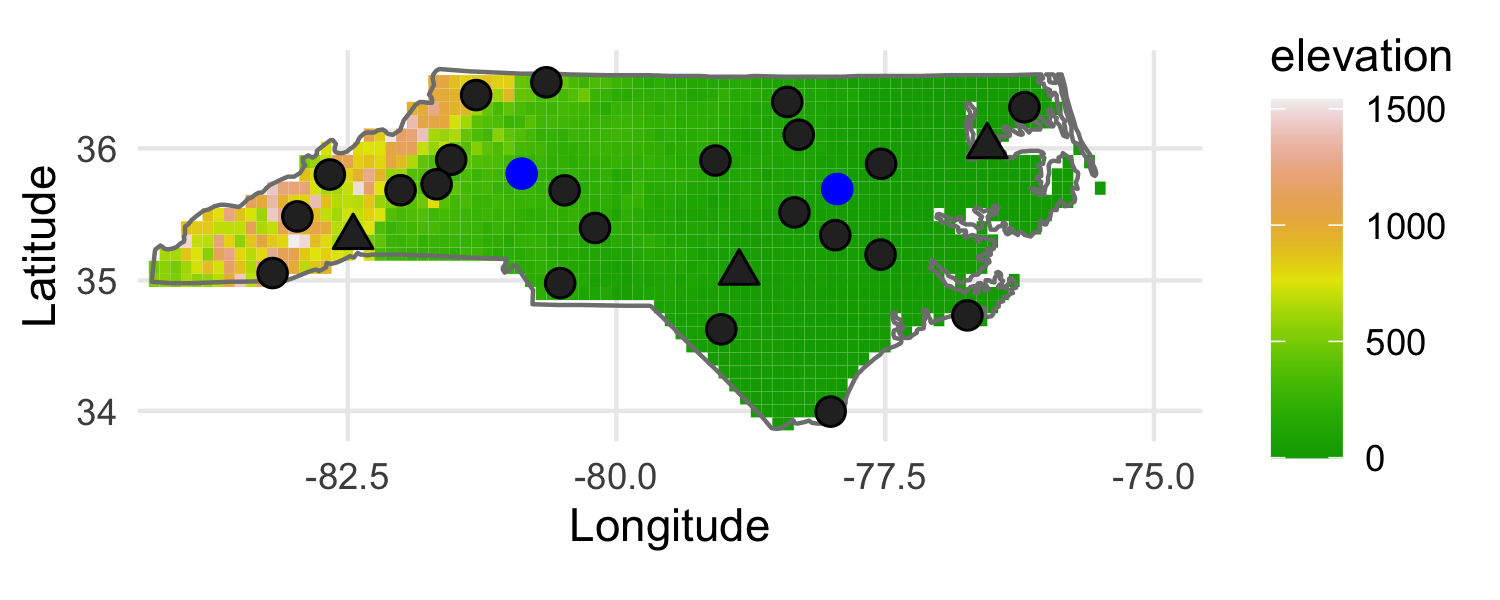}
	\caption{Map of North Carolina showing the sites and altitude in meters above sea level of the weather stations. The sites marked by black symbols were used to fit the model, the remaining to validate the model.}  
	\label{fig:stazioni_nc}
\end{figure} 

Convergence of  the MCMC chains has been assessed using several diagnostic techniques \citep{brooks_gelman}. Traceplots are reported in Figure S5 of the supplementary materials.
After establishing the convergence of the chains we assess whether the parametric assumptions of the proposed model provide a good fit to the observed data. We perform posterior predictive checks \citep{gelmanbda03}, comparing relevant quantities, such that $y_i$ or $x_{ij}$, with their corresponding posterior predictive densities. The posterior predictive distributions are not analytically available, but it is straightforward to simulate new data from them.
Examination of the posterior predictive distributions for the annual maxima, number of events, and daily rainfall magnitudes for two of the example stations are given in Figure S6, in the supplementary materials. Overall these posterior predictive distributions are satisfactorily captured by sHMEV, even if there are some discrepancies for the distribution of $n_j(s)$. However, as discussed in Section \ref{sHMEV_rain}, the distribution of $n_j(s)$ mainly affects the estimation of extreme events only through its average value, which appears to be adequately  captured by the model. Note that a discrepancy appears for small values of daily rainfall magnitudes, due to the sensitivity of the measurement instruments. However, the right tail of the daily precipitation distribution, which plays an important role in determining extreme values, is well captured by sHMEV.

The average  time for running the MCMC sampling for this dataset is below 18 minutes on a Apple M1 CPU laptop with 8 GB of RAM.

\subsection{Results}

A summary of the posterior distributions for $\bm{\beta_{\delta}}$, $\bm{\beta_{\gamma}}$, $\bm{\beta_{\lambda}}$, $\sigma_{\delta}$ and $\sigma_{\gamma}$ is given in Table \ref{tab:post_beta}. Overall, the distributions show low dispersion and the effect of the covariates on precipitation intensity and occurrence agrees with what was observed in the exploratory analysis.

\begin{table} [t]
\caption{
Summary statistics for the posterior distributions of the latent process parameters referred to $\gamma$ and $\delta$,  respectively the shape and scale parameters of the Weibull distribution and $\lambda$, the success probability of the binomial distribution for the number of events. The posterior means and the associated 95\% credible intervals (parentheses) are displayed. 
}
\label{tab:post_beta}
\begin{center}
\begin{tabular}{lccc}
\toprule
& $\gamma$ & $\delta$ & $\lambda$ \\
\midrule
 $\beta_0$ & $  0.86  \ (0.85, \ 0.87)$ & $10.5 \ (10.3, \ 10.7)$ & $-0.93 \ (-0.94, \ -0.92)$ \\
 $\beta_1$(lat) & $  0.02 \ (0.01, \ 0.03)$ & 
 $0.01 \ (-0.25,\ 0.26)$ & $0.14 \ (0.11,\  0.17)$ \\
 $\beta_2$(lon) & $  0.2 \ (-0.04, \ 0.07)$ & $0.49 \ (-0.21, \ 1.17)$ & $-0.7 \ (-0.85, \ -0.55)$ \\
 $\beta_3$(alt) &$-0.03 \ (-0.05,  \ -0.01)$ &
 $0.01 \ (-0.46, \ 0.45)$ & $0.13 \ (0.11,\  0.15)$ \\
 $\beta_4$(coast) & $ 0.01 \ (-0.06, \ 0.07)$ & $-0.28 \ (-1.02,\  0.43)$ & $-0.7 \ (-0.85, \ -0.55)$ \\
 $\sigma$ & $ 0.09 \ (0.08, \ 0.10)$ & $2.19 \ (2.01, \ 2.38)$ \\

\bottomrule
\end{tabular}
\end{center}

\end{table}

Our goal is to estimate the posterior distribution for the return level for every location in North Carolina. Since the cdf of annual maxima is a function of the parameter $\gamma$, $\delta$ and $\lambda$, it is sufficient to estimate the posteriors of these processes. We divide the study region into a grid of points and we consider the values of latitude, longitude, altitude and distance from the coast for each point.
With the posterior distributions of $\bm{\beta_{\delta}}$, $\bm{\beta_{\gamma}}$ and $\bm{\beta_{\lambda}}$ and the values of the covariates it is immediate to compute $\mu_{\gamma}$, $\mu_{\delta}$ and $\mu_{\lambda}$ for each point of the map.
Doing this for each iteration $b$, $b = 1 \dots B=4000$, provides draws from the posterior distribution of $\lambda$ and simulating from a Gumbel distribution we get also draws from the posterior distribution of $\gamma$ and $\delta$.

Figure S7 in the supplementary materials shows the pointwise mean and the pointwise interquartile range of the posterior draws over the $B$ iterations. The Weibull shape parameter, $\gamma$, is lower for the mountainous area and tends to increase as latitude increases.  For the scale parameter $\delta$, a positive effect of longitude and distance from the coast is observed. Finally, the map for $\lambda$ shows that the number of annual rainy days, as might be expected, is higher in mountainous areas and decreases along the coast.
The level of uncertainty for all three parameters is greatest at some mountain locations.

Figure \ref{fig:mappe_return} shows maps of the predictive pointwise posterior mean for the 25 and 50 year return levels, with pointwise 90\% credible intervals.  Adopting a Bayesian methodology, indeed, allows to obtain natural uncertainty estimates for the return levels, taking the pointwise 0.05 and 0.95 empirical quantiles from the return level draws. The return levels were calculated as described in Section \ref{sec:prior} using (\ref{quant_sHMEV2}) and (\ref{quant_sHMEV1}).
We observe higher values for the mountainous and southwest areas, where there are few stations and the model is forced to extrapolate. In general, there also appears to be a slightly decreasing trend with latitude. 

We want to evaluate the performance of the proposed model in predicting extreme values for sites not in the data, i.e., by extrapolating to space.
For this purpose, we calculate for several return times the quantiles for the two stations in the test set in space, which were not used to estimate the model. Figure \ref{fig:valid_spat} reports the quantile versus return time plots for the stations of Statesville and Wilson. For Statesville the quantiles predicted by the sHMEV model match observed annual maxima very well, for Wilson, instead, the model predictions turn out to be less good. 
However, for this site there are three rather high observed values that are quite difficult to predict. These values were observed during storms and cyclones that caused significant economic and social damages for Wilson county: the tropical storm Brenda on July 29, 1960, the hurricane Floyd on September 16, 1999 and the hurricane Matthew on October 8, 2016. Thus we are not dealing only with heavy rainfall, but with widespread phenomena that involve several extreme meteorological events and require a larger physical investigation.
We also recall that the model was estimated using only the first 20 years of the time series and exploiting all the available temporal information could lead to a more accurate spatial extrapolation.

\begin{figure}[t]
	\centering
    \includegraphics[width=0.9\textwidth]{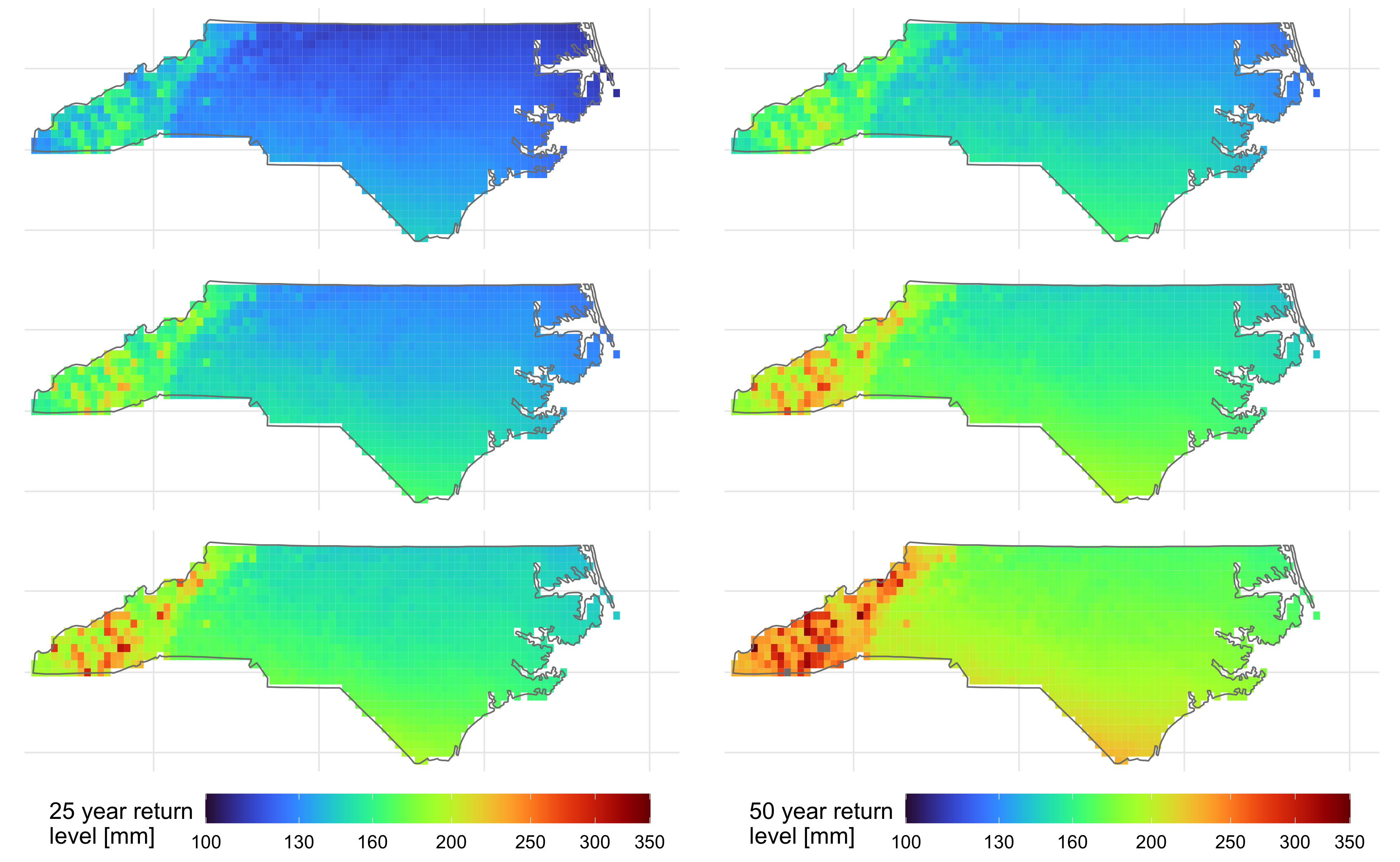}
	\caption{Maps of the predictive pointwise 25 and 50 year return level estimates for rainfall (mm) obtained from the sHMEV model. The top and bottom rows show the upper and lower bounds of the 90\% pointwise credible intervals, the middle rows show the predictive pointwise posterior mean. }  
	\label{fig:mappe_return}
\end{figure} 

\begin{figure}[t!]
	\centering
	\includegraphics[width=0.65\textwidth]{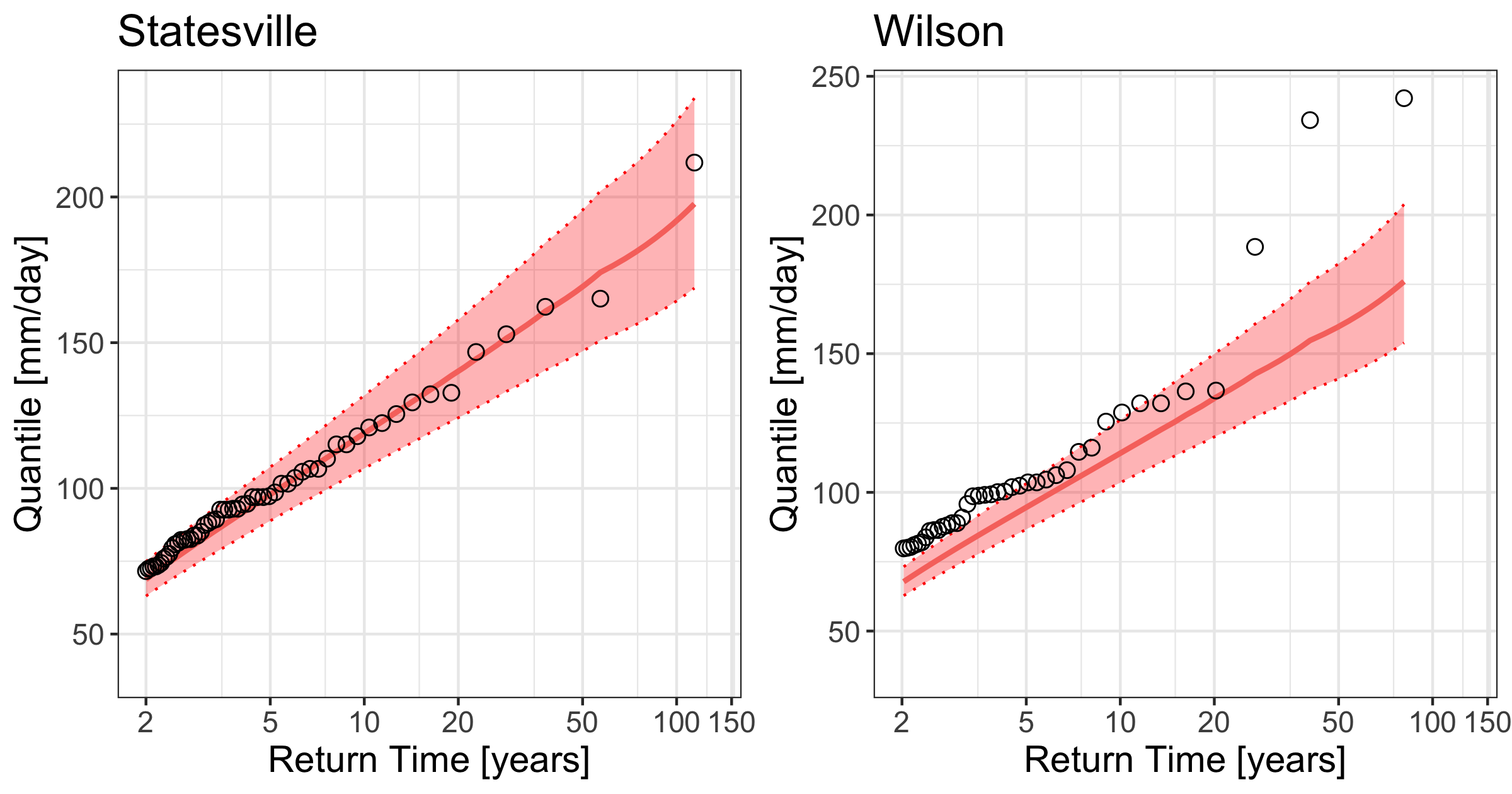}
	\caption{Quantiles predicted for the two stations (Statesville and Wilson) taken as the test set on the space. Solid lines show the expected value of the quantile for a given return time, while dashed lines represent the bounds of 90\% credibility intervals. Circles represent the observed maxima.}  
	\label{fig:valid_spat}
\end{figure}

In the following, we compare extreme value quantiles obtained from the HMEV, GEV, and sHMEV models, evaluating the predictive accuracy in estimating the empirical distribution of annual maxima in the test set over time. In fact, all three models were trained on just the first 20 years of record. 

For the GEV model the shape parameter $\tau$ in equation (\ref{eq:gev}) plays a crucial role, since it determines the behavior of the tails of the distribution. 
Similar to other dataset on extreme rainfall, the estimates of the shape parameter are positive, corresponding to the heavy-tailed Fréchet case, but not strongly so \citep{papax, padoan_davison}. Furthermore, these estimates agree with the ones obtained by estimating a GEV model on the data simulated from the posterior predictive distribution of the sHMEV model. 

Figure S8  in the supplementary materials shows the empirical distribution of the indexes FSE, $b_q$ and $\Delta_{q90}$ over the 27 stations. Regarding FSE, the sHMEV model provides the best results for a larger number of stations, however it has rather high values for some sites. Looking specifically at these sites, they are mainly located on the coast. Probably for these stations, which exhibit a specific behavior, the sharing of information among different sites results in an excessive shrinkage to the mean, leading to higher errors in the prediction of extreme values.
Regarding estimation bias, both sHMEV and HMEV tend to underestimate quantiles. As regards variability, the sHMEV model is the most efficient  and it provides considerably thinner credible intervals than the GEV model. This reduction in variability results in a small bias in point estimation, consistent with a well–known tradeoff in statistics known as the variance–bias tradeoff. However, when considering a global measure such as the fractional squared error the proposed approach still outperforms the alternatives.
These results are consistent with those obtained in the simulation study.

Figure S9 in the supplementary materials reports the quantile versus return time plots of the different competing methods for two of the stations taken as an example.
Models estimates differ, with sHMEV exhibiting, as previously observed from our simulation study, narrower credibility intervals with respect to HMEV and GEV models. 
For the Fayetteville station the sHMEV model presents an overall good agreement with the empirical frequencies associated with the annual maxima extracted from the entire record. The GEV model, instead, is more influenced by the specific training set  used and tends to overestimate the values.
For the Edenton station none of the three models adequately captures the distribution of the annual maxima. It should be noted that this station is on the coast, and is likely to experience very unusual and difficult to predict precipitation.

\section{Discussion}\label{sec5}

We introduced a spatial hierarchical Bayesian model to analyze environmental extreme values. The proposed approach, which extends and generalizes the hierarchical model of \citet{hmev}, avoids the asymptotic arguments of classical extreme values models and exploits most of the information contained in the data through the ordinary events. 
The spatial dependence has been induced in the parameters determining the ordinary events and, specifically, modeled through a linear combination of spatial covariates with unknown regression parameters. The Bayesian approach allowed the inclusion of valuable prior information that is often present in environmental modeling. The performance of the method are competitive with the state of the art methods and with the original proposal of  \citet{hmev} not exploiting any spatial information. 

While we focused on simple formulations for the parameters related to the ordinary events, more complex formulations are also possible including semiparametric specifications, for example exploiting spline regressions, or including random noise through suitable stochastic process modeling the residual spatial dependence, e.g. Gaussian processes. 

The inclusion of spatial covariates sheds light on possible extensions. For example, time dependence can be introduced through a suitable function of covariates or including trend and seasonality. One might also consider a regularized version of the method, by setting a suitable shrinkage prior on the regression coefficients, see \citet{Carvalho2022regulariz}.

%Where the bibliography will be printed
\nocite{*}
\printbibliography[heading=bibintoc]

\newpage
\section*{Supplementary materials}
\setcounter{figure}{0}
\setcounter{section}{0}
\renewcommand{\thefigure}{S\arabic{figure}}
\renewcommand{\thesection}{S\arabic{section}}

\section{Details on the Bayesian GEV and HMEV implemented}
Here we briefly described the methods used to benchmark the sHMEV model in the analysis: a Bayesian implementation of the classical generalized extreme value distribution (GEV) and the Bayesian hierarchical model (HMEV) described in \citet{hmev}.

For fitting the GEV model,  whose cdf is given in equation (1) in the paper, we use Bayesian methods. 
As done for the sHMEV model, we implemented a Stan model, sampling from the posterior using the Hamiltonian Monte Carlo approach. For the shape parameter $\tau$ we take a Normal prior centered in 0.114 and with standard deviation equal to 0.125. This choice is motivated by the expected value suggested globally for daily rainfall extremes \citep{papax}. For the location parameter $\mu$ we choose a Normal prior, while for the scale parameter $\sigma$ a gamma prior. We centered the Normal and the gamma priors around the mean and standard deviation of the annual maxima samples respectively.

For the HMEV model we follow the specific formulation for modeling daily rainfall described in \citet{hmev}. Denoting with $x_{ij}$ the magnitude of the i-$th$ event within the $j$-th block  and with $n_j$ the number of events observed over the $j-$th block ($i = 1, \dots,n, j=1, \dots, J$), they assumed the following hierarchical model
\begin{align*}
	n_{j } \mid \lambda \sim Binomial(N_{t}, \lambda), \quad \delta_{j} \mid \mu_{\delta}, \sigma_{\delta} \sim Gumbel(\mu_{\delta}, \sigma_{\delta}), \\
	 \quad \gamma_{j} \mid \mu_{\gamma}, \sigma_{\gamma} \sim Gumbel(\mu_{\gamma}, \sigma_{\gamma})
	 \quad x_{ij} \mid n_{j}, \gamma_{j}, \delta_{j} \sim Weibull(\gamma_{j}, \delta_{j}),
\end{align*}
with $N_{t} = 366$. The hierarchical representation of the model is completed by eliciting suitable distributions, that exploit physical knowledge on the rainfall phenomena. For the parameters $\mu_{\gamma}, \sigma_{\gamma}, \mu_{\delta}$ and $\sigma_{\delta}$ they opted for independent inverse gamma distributions, while for $\lambda$ they chose a beta distribution. We used a Stan implementation to estimate the posterior distributions. For more details see \citet{hmev}.

\newpage
\section{Additional plots}

\begin{figure} [h]
	\centering
		\includegraphics[width=0.99\textwidth]{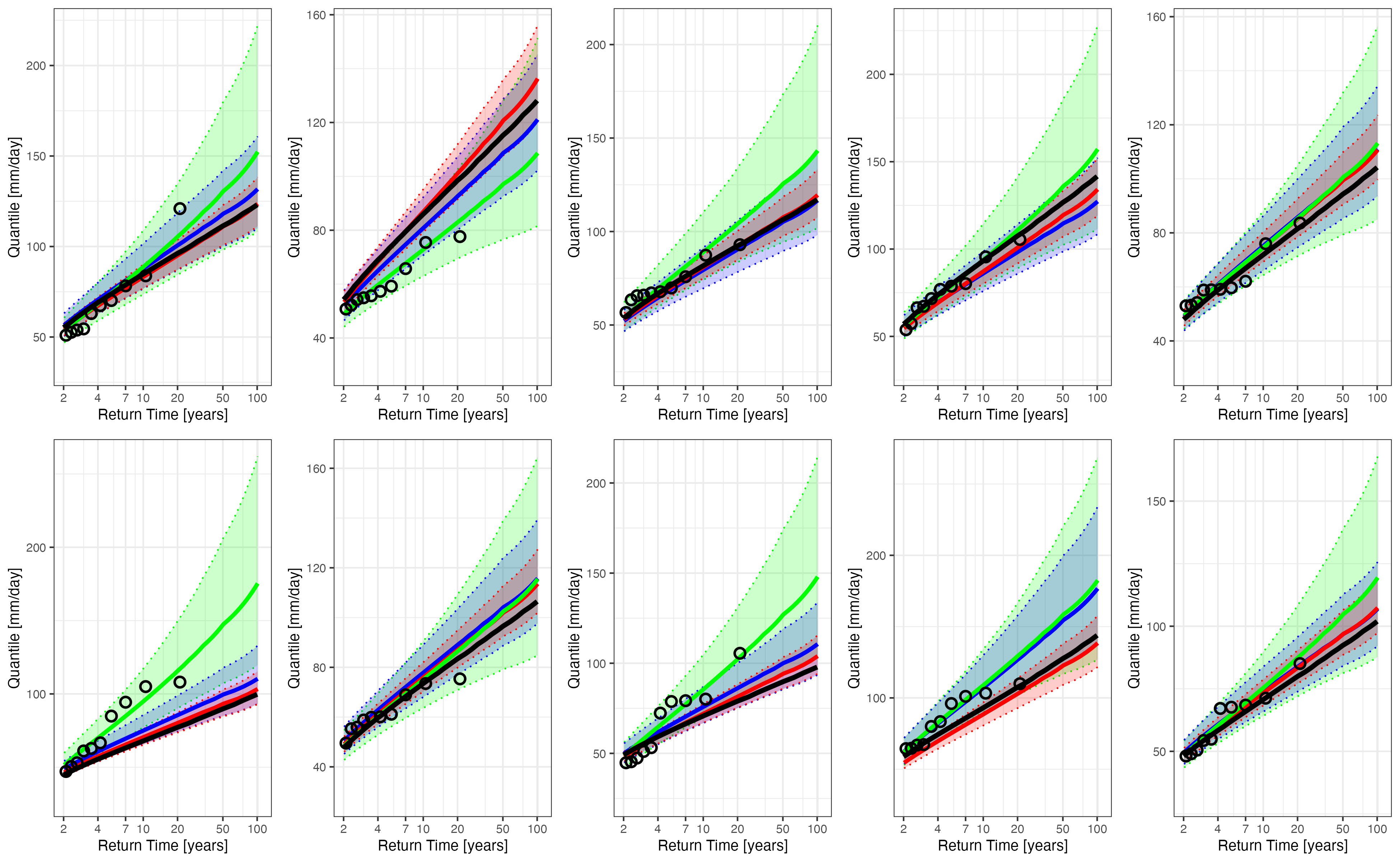}
	\caption{Quantiles predicted by the GEV (green), HMEV (blue), and sHMEV (red) models based on data simulated in the first scenario (WEI) for different stations. Solid lines show the expected value of the quantile for a given return time, while dashed lines represent the bounds of 90\% credibility intervals. Circles represent the observed block maxima on training set, while the black lines report the quantiles computed from the true sHMEV model.}  
	\label{fig:qrplot1}
\end{figure} 

\begin{figure} 
	\centering
		\includegraphics[width=0.99\textwidth]{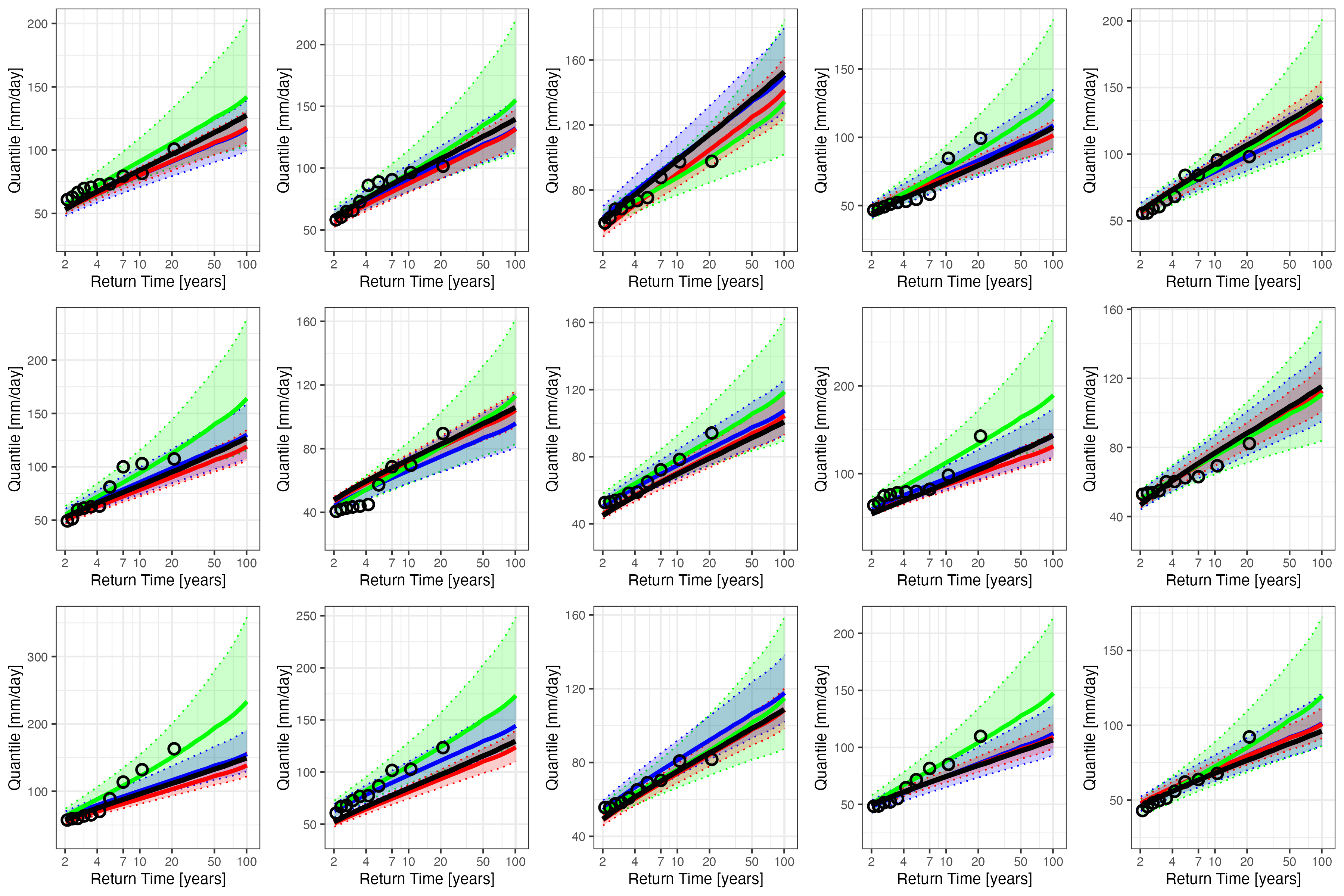}
	\caption{Quantiles predicted by the GEV (green), HMEV (blue), and sHMEV (red) models based on data simulated in the first scenario (WEI) for different stations. Solid lines show the expected value of the quantile for a given return time, while dashed lines represent the bounds of 90\% credibility intervals. Circles represent the observed block maxima on training set, while the black lines report the quantiles computed from the true sHMEV model.}  
	\label{fig:qrplot2}
\end{figure}

\begin{figure}
	\centering
		\includegraphics[height = 0.17\textheight]{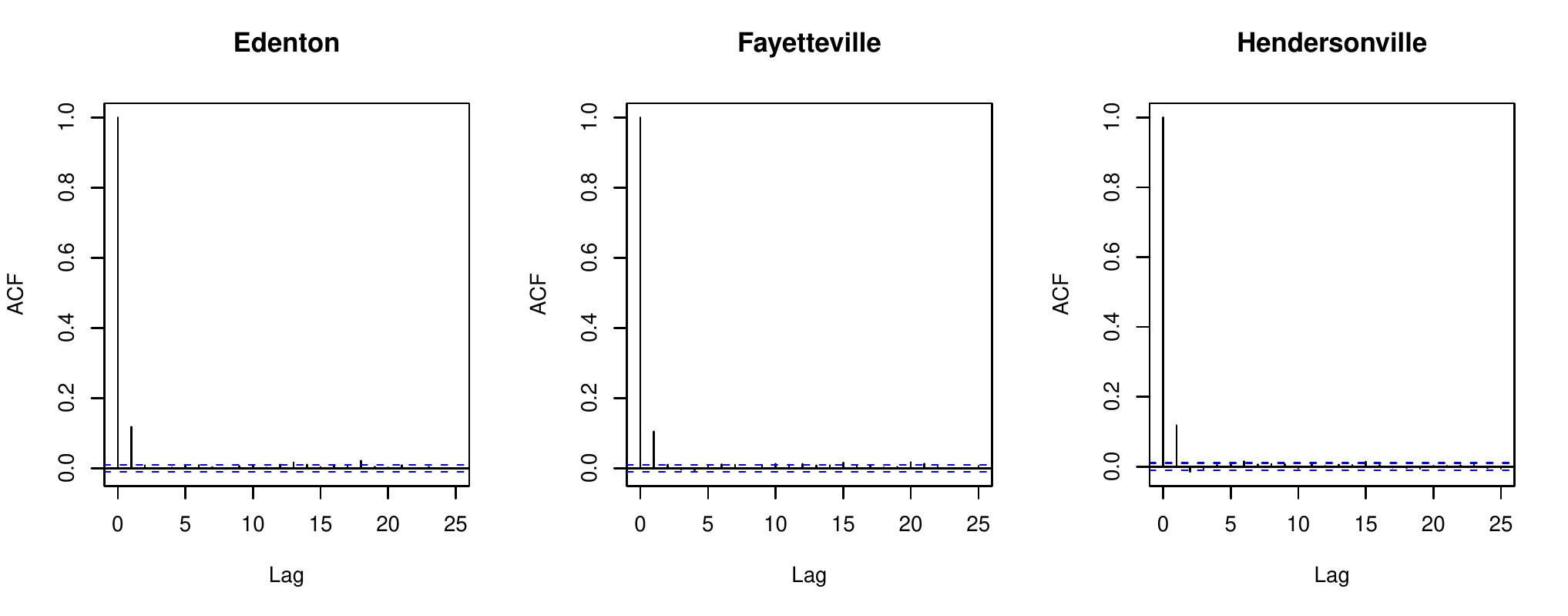}
	\caption{ACF of the time series for daily precipitation accumulation for Edenton, Fayetteville and Hendersonville stations.}  
	\label{fig:acf}
\end{figure} 

\begin{figure}
	\centering
		\includegraphics[height = 0.2\textheight]{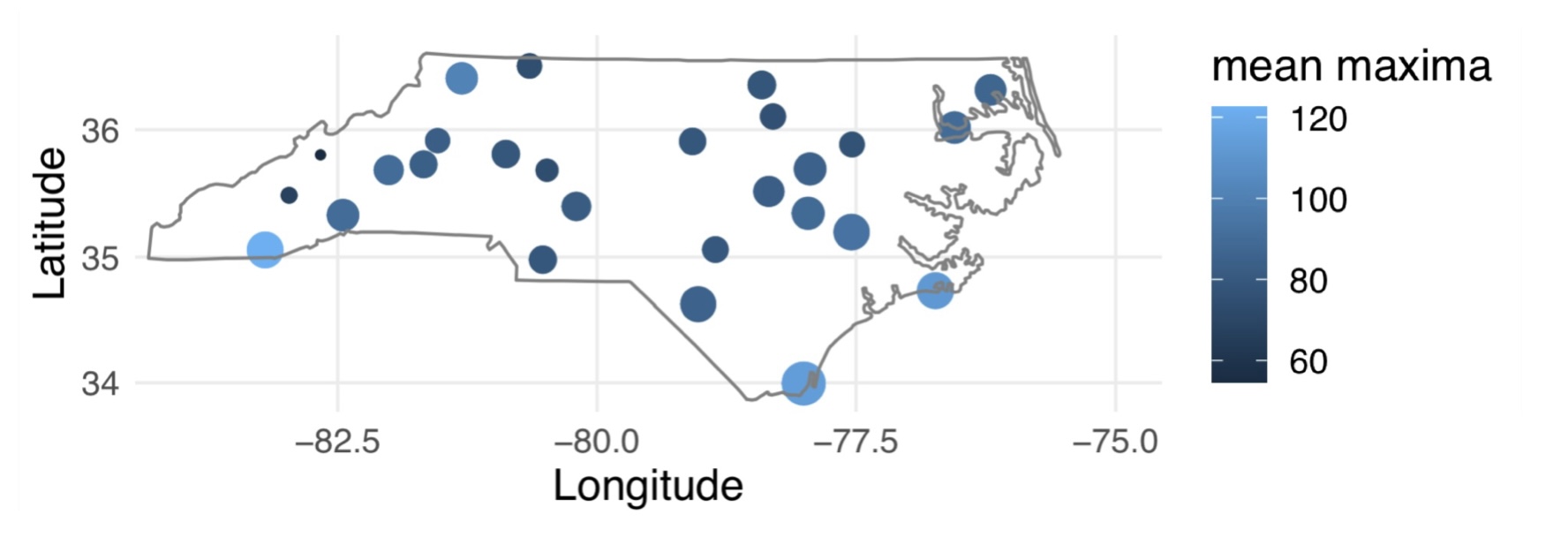}
	\caption{Annual maxima of daily rainfall accumulations for each station. The color and the size of the circles are respectively proportional to the mean and the standard error of the annual maxima.}  
	\label{fig:max}
\end{figure} 

\begin{figure}
	\centering
		\includegraphics[width=0.99\textwidth]{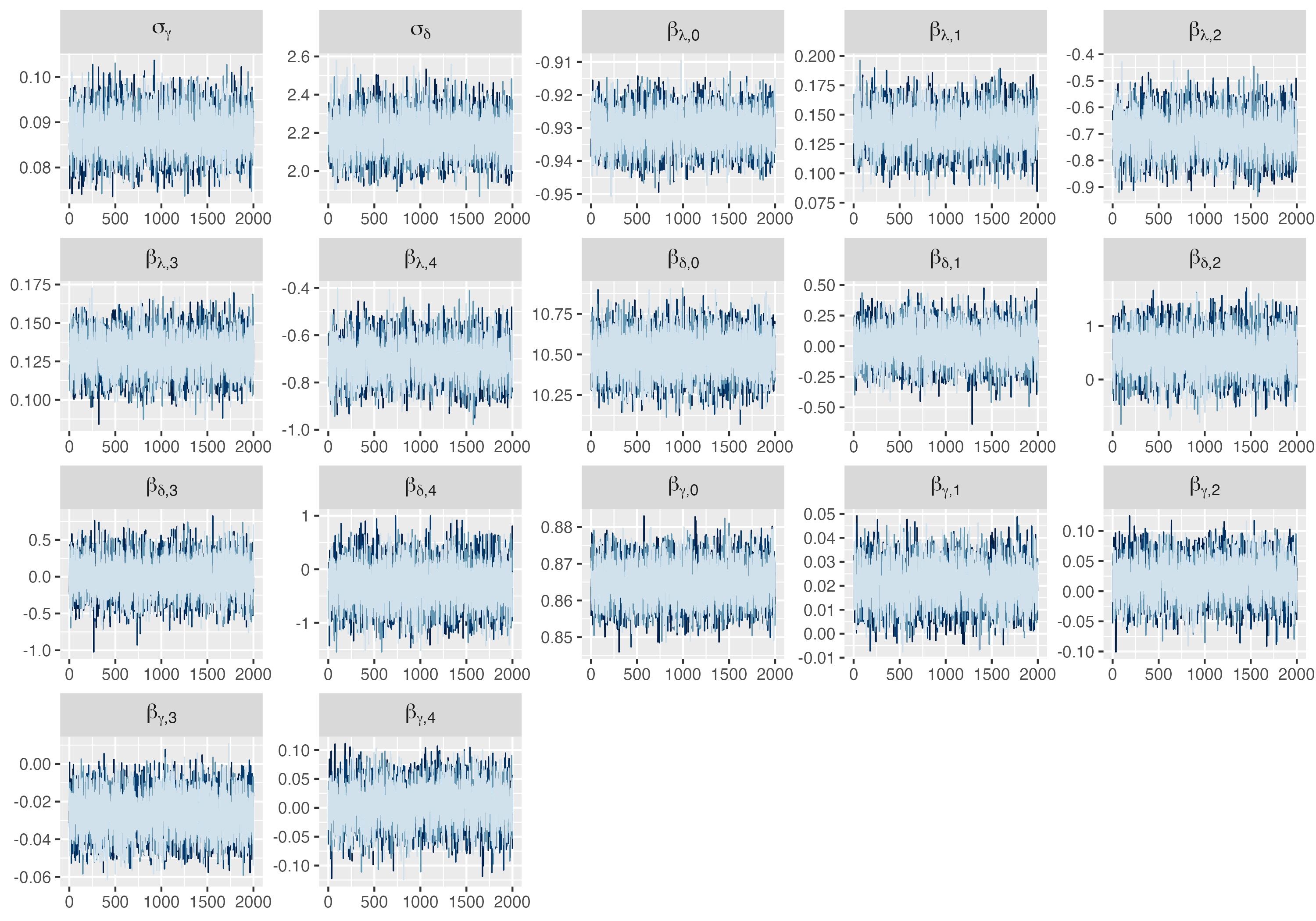}
	\caption{Trace plots, without burn-in, for each parameter for the sHMEV model estimated on the North Carolina rainfall data. Different colors represent the 4 chains simulated in parallel..}  
	\label{fig:trace}
\end{figure}

\begin{figure} 
	\centering
	\includegraphics[width=0.6\textwidth]{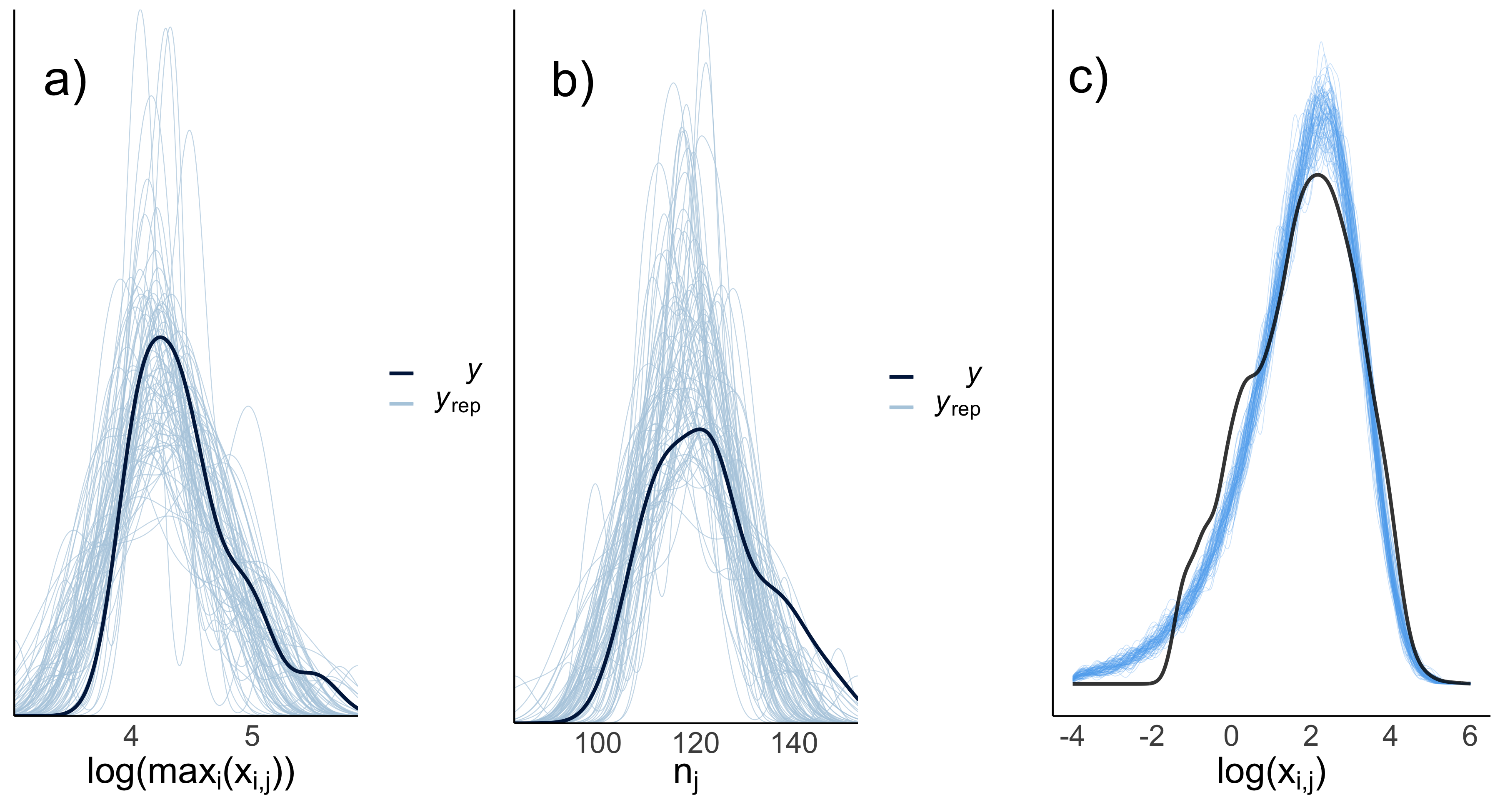}
	\includegraphics[width=0.6\textwidth]{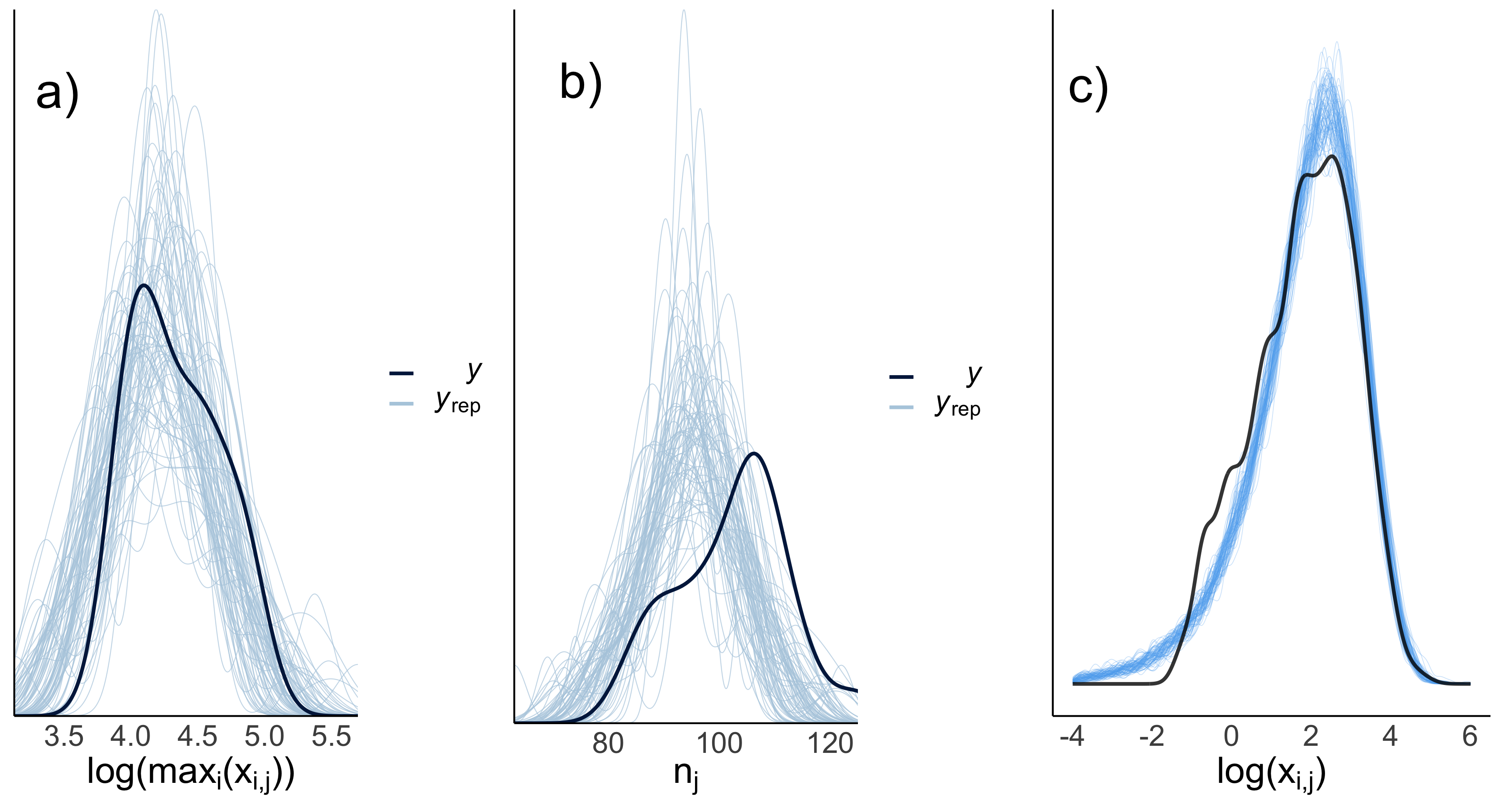}
	\caption{Posterior predictive distributions for the logarithm of the annual maximum daily rainfall accumulations (a), yearly number of events (b) and logarithm of non-zero daily rainfall events (c). The  top rows show the results for the station of Hendersonville, the bottom ones for the station of Fayetteville.   Dark blue lines show the density of the observed values (obtained by kernel density estimation), while the light blue lines show the kernel density estimates for 100 draws from the posterior predictive distributions.}  
	\label{fig:ppd}
\end{figure}

\begin{figure}
	\centering
	\includegraphics[width=0.98\textwidth]{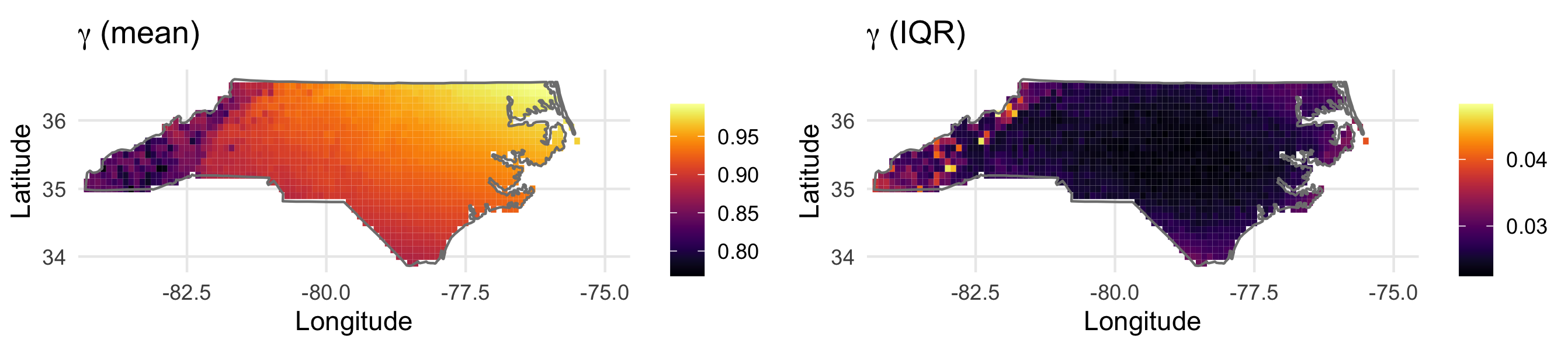}
    \includegraphics[width=0.98\textwidth]{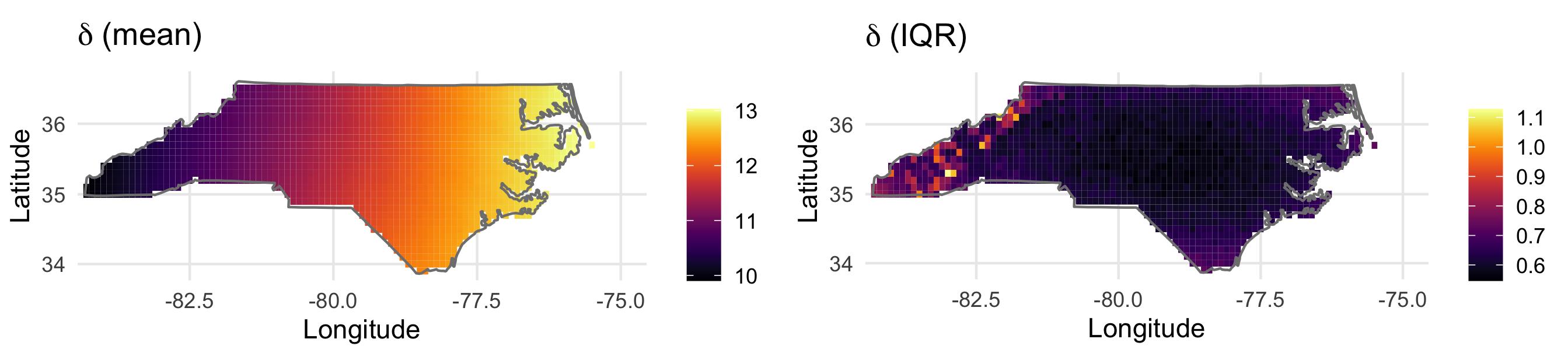}
    \includegraphics[width=0.98\textwidth]{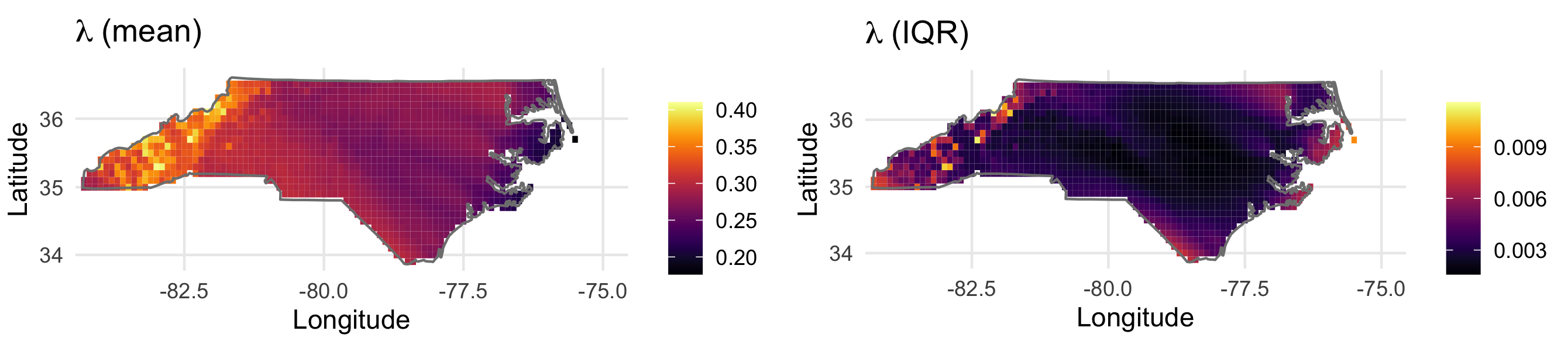}
	\caption{Maps with posterior predictive distributions for the Weibull parameters, $\delta$ and $\gamma$, and for the success probability of the binomial distribution, $\lambda$. The figures show the posterior mean (on the left) and the interquartile range (on the right) of the distributions.}  
	\label{fig:mappe_post}
\end{figure}

\begin{figure} 
	\centering
	\includegraphics[height = 0.24\textheight]{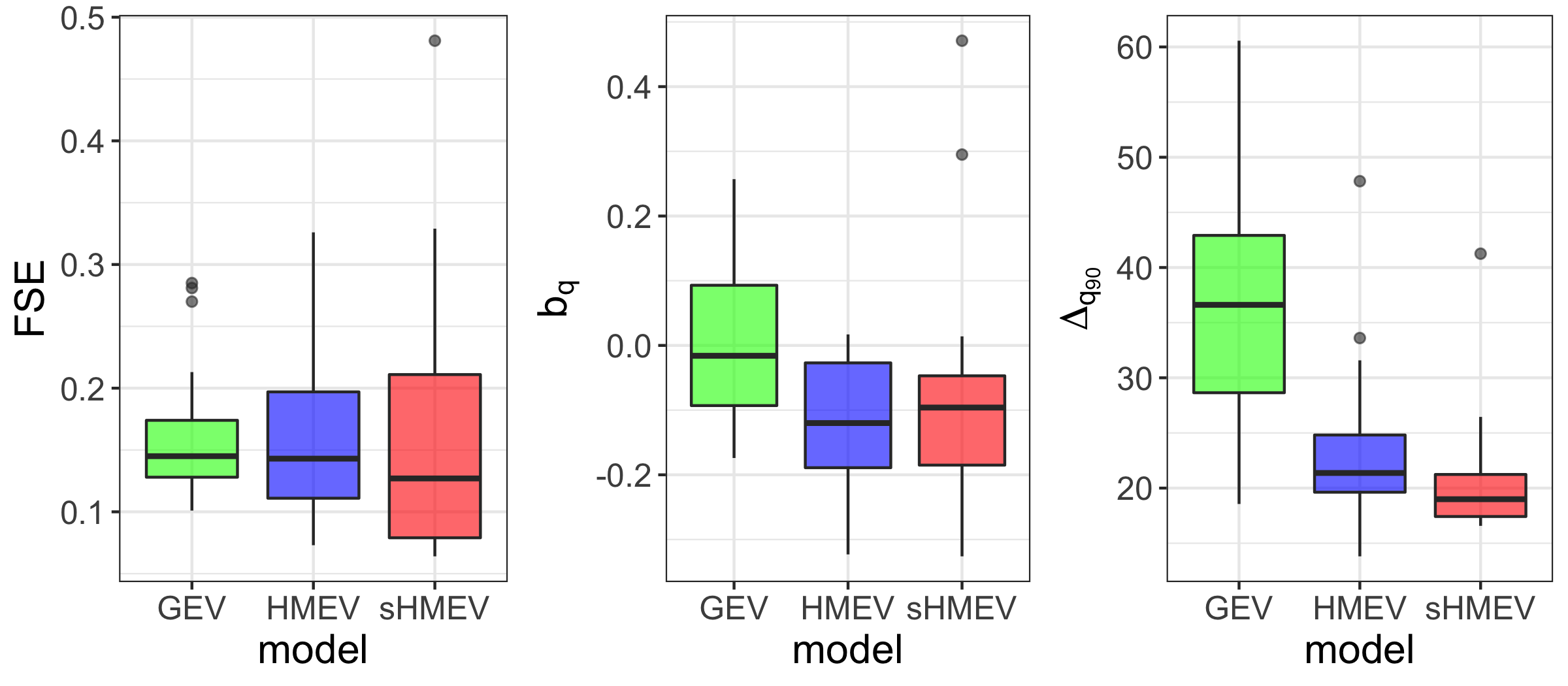}
	\caption{FSE, mean bias and mean credibility interval width.}  
	\label{fig:fse_nc}
\end{figure}

\begin{figure}[!t]
	\centering
	\includegraphics[width=10cm]{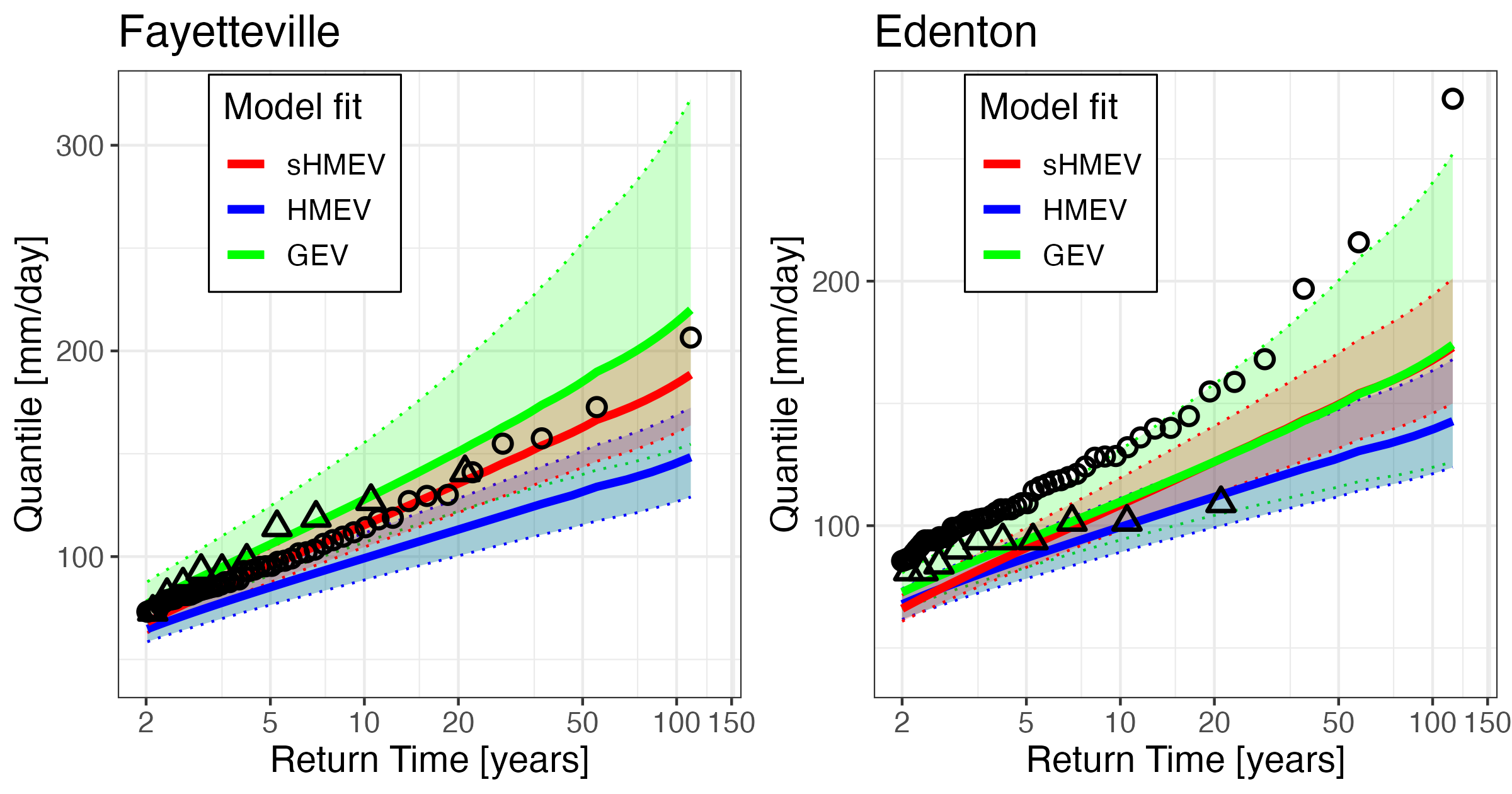}
	\caption{Quantiles predicted for the stations of Fayetteville and Edenton by the GEV (green), HMEV (blue), and sHMEV (red) models. Solid lines show the expected value of the quantile for a given return time, while dashed lines represent the bounds of 90\% credibility intervals. Triangles represent the maxima on the training set, while the circles represent the maxima on the test set.}  
	\label{fig:quant_compare}
\end{figure} 
\end{document}